\documentclass[twocolumn, twocolappendix]{aastex63}

\usepackage{comment}

\usepackage[italicdiff]{physics} 
\usepackage{amsmath}

\usepackage{graphicx}
\usepackage{float}

\usepackage{bigstrut}

\usepackage{tabularx}
\newcolumntype{C}{>{\centering\arraybackslash}X}

\newcommand{\m}[1]{\underline{\underline{#1}}}

\newcommand{\avM}[1]{\left\langle #1 \right\rangle}  
\newcommand{\avV}[1]{\left\langle #1 \right\rangle_{V}}  

\newcommand{\hi}{\textsc{H\,i}}
\newcommand{\hii}{\textsc{H\,ii}}

\usepackage[encapsulated]{CJK}
\usepackage{ucs}
\usepackage{url}
\newcommand{\cntext}[1]{\begin{CJK}{UTF8}{bsmi}#1\end{CJK}}

\defcitealias{Micic2012}{M12}

\shorttitle{Mixing vs. Chemistry}

\begin{document}


\title{Clumping Factor in the Mix: Incomplete Mixing Suppresses Turbulent Reaction Boost}

\author{Yu-hsuan Liu (\cntext{劉宇軒})}
\affiliation{Institute of Astrophysics \& Department of Physics, National Taiwan University, No. 1, Sec. 4, Roosevelt Rd., Taipei 10617, Taiwan}

\author[0000-0002-9235-3529]{Chia-Yu Hu (\cntext{胡家瑜})}
\affiliation{Institute of Astrophysics \& Department of Physics, National Taiwan University, No. 1, Sec. 4, Roosevelt Rd., Taipei 10617, Taiwan}
\affiliation{Physics Division, National Center for Theoretical Sciences, Taipei 10617, Taiwan}

\correspondingauthor{Chia-Yu Hu}
\email{huchiayu@phys.ntu.edu.tw}

\begin{abstract}

Cosmological simulations often account for unresolved density fluctuations using a sub-grid clumping factor when modeling chemical reactions.
To validate this approach,
we conduct hydrodynamical simulations of isothermal, supersonic turbulence and track the evolution of passive chemical species in first-, second-, and third-order reactions,
exploring seven reaction rates across Mach numbers $\mathcal{M} = 2$ -- 50 in three turbulence driving modes.
We parametrize the saturation of clumping factors at $\mathcal{M} \gtrsim 10$ due to intermittency (i.e., falling significantly below the log-normal prediction) with a one-parameter fitting formula.
Furthermore,
we find that the competition between turbulent mixing and the density-dependent reaction rates,
governed by the Damk\"ohler number (Da),
suppresses the global reaction rate by a \textit{mixing efficiency} factor ($\xi$).
For Da $\ll 1$, chemical species remain well mixed ($\xi\to 1$), and the reaction rates are boosted by a clumping factor.
For Da $\gg 1$, reactants in dense gas are rapidly depleted before being redistributed by turbulence, causing $\xi$ to drop by a factor of a few to more than an order of magnitude.
We quantify the effective mixing efficiency as a function of two dimensionless parameters, $\mathcal{M}$ and Da, and provide a fitting formula that can be easily adopted in cosmological simulations to accurately follow unresolved chemical reactions.

\end{abstract}
\keywords{hydrodynamical simulations; interstellar medium; astrochemistry}

\section{Introduction}
\label{sec:introduction}

Chemistry plays a crucial role in the evolution of the interstellar medium (ISM) \citep{Tielens2005, Draine2011, Klessen2016}.
Chemical reactions influence the thermal balance of gas through radiative cooling via fine-structure metal lines and rotational/vibrational molecular lines.
Surface reactions on dust grains lead to the formation of molecular hydrogen (H$_2$),
the primary constituent of molecular clouds where stars are born,
which ultimately triggers the gas cycle in the ISM and regulates galaxy evolution. 
Observationally,
different chemical species provide unique signatures through spectral lines, providing valuable diagnostics of the physical conditions such as gas density, temperature, and the ambient radiation field.

Most of these reactions have rates that depend on local gas density ($\rho$) due to their collisional nature.
For example,
two-body collision processes such as H$_2$ formation on dust, collisional ionization, recombination, and most gas-phase chemical reactions have (volumetric) reaction rates that scale as $\rho^2$.
Three-body reaction rates, such as gas-phase H$_2$ formation ($\mathrm{3 H \to H_2 + H}$), scale as $\rho^3$, though they are typically subdominant due to the extremely low densities in the ISM.
However, they play a critical role in forming the first stars in the early universe where dust is absent \citep{Abel2002, Bovino2014, Dutta2015}.
In contrast,
the temperature variation in molecular clouds is much weaker than the density variation, so clouds can largely be approximated as isothermal \citep{Wolfire2003, Larson2005, Heyer2015}.
The density structure of molecular clouds is therefore the most crucial information for modeling chemistry.

Observations of molecular linewidths have long established that molecular clouds are supersonically turbulent, with typical Mach numbers ranging from $\mathcal{M}\sim$ 5--50 \citep{Zuckerman1974, Larson1981, Heyer2004, RomanDuval2011, Heyer2015}.
Theoretical models and numerical simulations have demonstrated that an isothermal, supersonically turbulent medium develops a density structure that follows a log-normal probability distribution function (PDF) \citep{Vazquez-Semadeni1994, Padoan1997, Passot1998}.
The variance of the log-normal function is dictated by $\mathcal{M}$ as
\begin{equation}\label{eq:sigma_s}
    \sigma_s^2 = \ln{(1 + b^2 \mathcal{M}^2)},
\end{equation}
where $b$ is the forcing parameter ranging from $1/3$ for purely solenoidal driving 
to $1$ for purely compressive driving \citep{Federrath2008, Federrath2010}.
This simple analytic form has led to significant development in star formation theory.
Specifically, it allows the star formation efficiency to be derived analytically,
though competing theoretical models differ in the exact critical density required for the onset of star formation \citep{Krumholz2005, Padoan2011, Hennebelle2011}.

However,
in large-scale cosmological or isolated galaxy simulations,
the internal structures of these molecular clouds remain severely unresolved,
where a single resolution element represents the entire cloud or a significant fraction of the cloud.
As such,
gas densities in these clouds can be substantially underestimated.
Several studies have implemented star formation sub-grid models based on the log-normal theory 
to account for the unresolved density structure \citep{Semenov2016, Kretschmer2020, Girma2024},
but the uncertainty remains in the choice of the critical density.
In contrast, modeling chemical reactions is more straightforward as it does not require choosing a critical density.
Therefore, the natural approach is to simply multiply the reaction rates by a so-called \textit{clumping factor}:
\begin{equation}
    C_2 \equiv
    \frac{\avV{\rho^2}}{\avV{\rho}^2},
\end{equation}
where the brackets denote the volume-weighted average of the sub-grid density PDF within a resolution element.
This factor accounts for the fact that local reaction rates are significantly enhanced in sub-grid, high-density clumps relative to the rate evaluated at the mean density of the resolution element.
For a log-normal density PDF,
the clumping factor is uniquely determined by $\mathcal{M}$.
Several simulations have adopted this framework, either by assuming a constant clumping factor \citep{Gnedin2009, Kuhlen2012, Christensen2012, Pallottini2017} or by calculating a spatially varying, $\mathcal{M}$-dependent clumping factor from the local velocity dispersion \citep{Lupi2018, Hopkins2023}. 

However,
while this approach accounts for the sub-grid density structures,
it remains incomplete.
\citet{Micic2012} (hereafter \citetalias{Micic2012}) investigated H$_2$ formation in a simulated turbulent medium with non-equilibrium cooling and chemistry, demonstrating that the \textit{effective clumping factor},
defined as the overall reaction boost compared to the uniform-density rate,
is significantly smaller than the conventional clumping factor.
This discrepancy arises because the conventional definition implicitly assumes rapid turbulent mixing.
However, once the high-density gas becomes fully molecular,
the reaction rate drops significantly.
The effect is stronger for compressive driving, where the difference can reach an order of magnitude.

Despite the caveat pointed out by \citetalias{Micic2012},
it is still unclear how to improve the model.
As such,
more than a decade after their work,
the conventional clumping factor remains the standard approach for unresolved reactions in cosmological simulations.
The main difficulty is that the effective clumping factor in \citetalias{Micic2012} is not universal and is expected to depend on local ISM conditions such as density, turbulent velocity, metallicity, cosmic ray ionization rate, etc.
Quantifying the effective clumping factor as a function of all relevant parameters is computationally formidable and requires a high-dimensional grid of turbulence simulations.

This paper aims to address the current limitation.
To this end,
we conduct hydrodynamical simulations of an isothermal, supersonically turbulent medium where chemistry is governed by a single, dominant process --- a good approximation for clouds sufficiently shielded from external radiation.
This makes the system universal, with the global reaction rate dictated by a competition between turbulent mixing and the local, density-dependent reaction rate.
Our simulations follow three common types of reactions in ISM chemistry separately,
and we show that the global reaction rate is fully characterized by two dimensionless parameters:
the turbulent Mach number ($\mathcal{M}$) and the \textit{Damk\"ohler number} (Da), defined as the ratio of the turbulent timescale to the chemical timescale\footnote{The use of the Damk\"ohler number was recently introduced to the astrophysical literature by \citet{Tan2021} in a different context (radiative turbulent mixing layers), where the role of chemical reaction is replaced by radiative cooling. Our usage is closer to the original definition in combustion.}.
We provide a parametrized fitting formula for use in large-scale cosmological simulations.

This paper is organized as follows.
Section~\ref{sec:theory} provides a theoretical framework for the three reaction types investigated in our simulations.
Section~\ref{sec:method} describes our simulation setup and the parameter space we explore.
Section~\ref{sec:results} presents our simulations, which quantify the clumping factor and mixing efficiency. We then parametrize these into ready-to-use sub-grid models.
Section~\ref{sec:discussion} discusses the implications of our results and compares them with previous work.
Section~\ref{sec:summary} summarizes our main findings.


\section{Analytic Theory}\label{sec:theory}

In this section, we briefly overview the three types of reactions commonly encountered in astrophysics.

In the interstellar medium,
reactions are dominated by two-body processes due to the extremely low gas densities.
The reaction rate depends on the product of the two reactants' densities multiplied by a rate coefficient.
The primary examples are \textit{hydrogen recombination} with free electrons in an \hii~region or \textit{collisional dissociation} of the molecular hydrogen (H$_2$).
In this case, the two reactants are depleted at a similar rate, and the rate equation can be expressed as
\begin{equation}\label{eq:xdot_2nd}
\dot{x} = -\frac{x^2}{\tau_2},
\end{equation}
with $\tau_2 = {R_2}/{\rho} \propto \rho^{-1}$ is the chemical reaction time,
$R_2$ is the rate coefficient,
and $\rho$ is the gas density.
Reactions that follow Eq.~\ref{eq:xdot_2nd} are called \textit{second-order} reactions, which have an analytic solution when evolved from $t_0$ to $t_1$:
\begin{equation}\label{eq:xsol_2nd}
x_1 = \frac{1}{1/x_0 + \Delta t/\tau_2},
\end{equation}
where 
$x_0 = x(t_0)$,
$x_1 = x(t_1)$,
and
$\Delta t = t_1 - t_0$.

In contrast, if one of the reactants is significantly more abundant than the other, the abundance of the former is essentially constant and thus can be absorbed into the rate coefficient, and the reaction rate depends only on the abundance of the latter:
\begin{equation}\label{eq:xdot_1st}
\dot{x} = -\frac{x}{\tau_1},
\end{equation}
where $\tau_1 = {R_1}/{\rho}$
and $R_1$ is the rate coefficient.
These are called the \textit{first-order} reactions with an analytic solution:
\begin{equation}\label{eq:xsol_1st}
x_1 = x_0\,e^{-\Delta t/\tau_1}.
\end{equation}
A primary example is the \textit{formation of molecular hydrogen} (H$_2$) on the surfaces of dust grains,
where the abundance of atomic hydrogen (\hi) is $x$
while the abundance of H$_2$ is $(1-x)/2$.
The dust-to-gas ratio is absorbed into the rate coefficient $R_1$.

In situations where densities are extremely high,
three-body processes can actually become important.
A primary example is the 3-body H$_2$ formation in the gas phase (3H $\rightarrow$ H$_2$), which plays an important role in primordial gas in the early universe.
The rate equation can be expressed as
\begin{equation}\label{eq:xdot_3rd}
\dot{x} = -\frac{x^3}{\tau_3},
\end{equation}
where
$\tau_3 = R_3/\rho^2$
and $R_3$ is the rate coefficient.
Eq.~\ref{eq:xdot_3rd} has an analytic solution 
\begin{equation}\label{eq:xsol_3rd}
x_1 = \frac{1}{\sqrt{\,x_0^{-2} + 2\Delta t/\tau_3\,}}.
\end{equation}

\subsection{Clumping Factor with Incomplete Mixing}

The reaction rates in Eqs.~\ref{eq:xdot_2nd}, \ref{eq:xdot_1st}, and \ref{eq:xdot_3rd} all depend on gas density due to their collision nature.
In a uniform medium,
these equations can be directly applied to describe the evolution of the global abundance, defined as $\avM{x} \equiv M_X / M$,
where $M$ is the total gas mass of the system and $M_X$ is the mass in the form of species $X$.
However, in an inhomogeneous medium, the reaction rate varies with the local density, which naturally leads to spatially varying abundances. 
The global abundance can thus be written as the \textit{mass-weighted} average abundance
\begin{equation}
    \avM{x} = \frac{\int x dM}{\int dM}.
\end{equation}


For the 1st-order reaction, if we take the mass-weighted average on both sides of Eq.~\ref{eq:xdot_1st}, we obtain
\begin{equation}\label{eq:av1st}
    \avM{\dot{x}} = -\frac{1}{R_1}\avM{\rho x}.
\end{equation}
However,
from the perspective of a sub-grid model, if we view the entire system as an unresolved resolution element, we do not have the information of $\avM{\rho x}$.
Instead, the only information we have is the mass-weighted average abundance $\avM{x}$ and the \textit{volume-weighted}\footnote{We denote $\avM{Q}$ (without an underscript) for the mass-weighted average of quantity $Q$ and $\avV{Q}$ (with an underscript $V$) for the volume-weighted average throughout this work.} average density, defined as
\begin{equation}
    \avV{\rho} \equiv  \frac{\int \rho dV}{\int dV} = \frac{M}{V}
\end{equation}
where $V$ is the total volume of the system.
As such, we can write Eq.~\ref{eq:av1st} as
\begin{align}\label{eq:xi1st}
\avM{\dot{x}}
&= -\frac{1}{R_1}\,\avM{\rho x} \nonumber\\
&= -\frac{1}{R_1}\,
\frac{\avM{\rho x}}{\avM{\rho}\avM{x}}\,
\frac{\avM{\rho}}{\avV{\rho}}\,
\avV{\rho}\avM{x} \nonumber\\
&= -\frac{1}{R_1}\,\xi_1 C_2\,\avV{\rho}\avM{x} \nonumber\\
&= -\frac{\xi_1 C_2}{\tau_1\!\left(\avV{\rho}\right)}\,\avM{x},
\end{align}
where
\begin{align}
    C_2 \equiv \frac{\avM{\rho}}{\avV{\rho}} 
    = \frac{\int \rho^2 dV / \int \rho dV}{\int \rho dV / \int dV} 
    = \frac{\avV{\rho^2}}{\avV{\rho}^2}
\end{align}
is the conventional \textit{two-body clumping factor} (note that $C_2\geq 1$)
and 
\begin{equation}
\xi_1 \equiv \frac{\avM{\rho x}}{\avM{\rho}\avM{x}},
\end{equation}
is what we refer to as the \textit{mixing efficiency}, which reflects the correlation between $\rho$ and $x$.
When turbulent mixing is efficient, the abundance is essentially constant everywhere (well-mixed), and thus $\xi_1 = 1$.
In contrast, if the reaction occurs rapidly compared to turbulent mixing, its density-dependent nature would naturally create non-uniform abundances,
with lower abundances at high densities due to faster depletion.
As such, $\rho$ and $x$ become anti-correlated, leading to $\xi_1 < 1$.




The same formulation can be applied to the 2nd-order reaction:
\begin{align}\label{eq:xi2nd}
\avM{\dot{x}}
&= -\frac{1}{R_2}\,\avM{\rho x^2} \nonumber\\
&= -\frac{1}{R_2}\,
\frac{\avM{\rho x^2}}{\avM{\rho}\avM{x}^{2}}\,
\frac{\avM{\rho}}{\avV{\rho}}\,
\avV{\rho}\avM{x}^{2} \nonumber\\
&= -\frac{\xi_2 C_2}{\tau_2\!\left(\avV{\rho}\right)}\,\avM{x}^{2},
\end{align}
where
\begin{equation}
\xi_2 \equiv \frac{\avM{\rho x^2}}{\avM{\rho}\avM{x}^{2}}.
\end{equation}
The clumping factor $C_2$ is the same as in the 1st-order equation, since both reactions originate from two-body collisions.
Similarly, the 3rd order equation can be expressed as
\begin{align}\label{eq:xi3rd}
\avM{\dot{x}}
&= -\frac{1}{R_3}\,\avM{\rho^2 x^3} \nonumber\\
&= -\frac{1}{R_3}\,
\frac{\avM{\rho^2 x^3}}{\avM{\rho^2}\avM{x}^{3}}\,
\frac{\avM{\rho^2}}{\avV{\rho}^{\,2}}\,
\avV{\rho}^{\,2}\avM{x}^{3} \nonumber\\
&= -\frac{\xi_3 C_3}{\tau_3\!\left(\avV{\rho}\right)}\,\avM{x}^{3},
\end{align}
where
\begin{equation}
\xi_3 \equiv \frac{\avM{\rho^2 x^3}}{\avM{\rho^2}\avM{x}^{3}},
\end{equation}
and
\begin{equation}
C_3 \equiv \frac{\avM{\rho^2}}{\avV{\rho}^{\,2}}
= \frac{\int \rho^3 dV / \int \rho dV}{(\int \rho dV / \int dV)^2} 
= \frac{\avV{\rho^3}}{\avV{\rho}^{\,3}} 
\end{equation}
is the \textit{three-body clumping factor} (note that $C_3\geq 1$).

Eqs.~\ref{eq:xi1st}, \ref{eq:xi2nd}, and \ref{eq:xi3rd} resemble the rate equations in a uniform medium at a density of $\avV{\rho}$ boosted by a factor of $\xi_1 C_2$, $\xi_2 C_2$, and $\xi_3 C_3$, respectively.
They reduce to the commonly adopted sub-grid model for unresolved chemical reactions if we set 
\begin{enumerate}
    \item 
    $\xi_1 = \xi_2 = \xi_3 = 1$, 
    \item 
    $C_2 = 1 + b^2 \mathcal{M}^2$, $C_3 = (1+b^2\mathcal{M}^2)^3$,
\end{enumerate}
where $b$ is the forcing parameter and $\mathcal{M}$ is the turbulent Mach number.
While the assumptions on $C_2$ and $C_3$ are well-established (though they require correction at high $\mathcal{M}$ as will be shown below),
the first assumption implicitly assumes that the reactant abundances are always perfectly mixed.
However, if the mixing efficiency is not 100\%, the reaction boost from turbulent clumping can be strongly overestimated.
The main goal of this paper is to systematically quantify the mixing efficiency and provide a more accurate sub-grid model that accounts for incomplete mixing.


\section{Numerical Methods}\label{sec:method}

Our turbulence simulations are conducted using {\sc Gizmo} \citep{Hopkins2015},
a public code for numerical hydrodynamics derived from the {\sc Gadget-2} code \citep{Springel2005}. 
We adopt the meshless finite-mass (MFM) solver \citep{Hopkins2015},
a strictly Lagrangian variation of the meshless Godunov method \citep{Gaburov2011}.
We assume a \textit{nearly} isothermal equation of state using a polytropic index $\gamma = 1.0001$.
Self-gravity is not included.

The simulation domain is a cube with a side length of $L = 1$.
The gas is initially uniform with a density $\rho = 1$ and the sound speed $c_{\rm s}$,
represented by $N^3$ gas particles arranged in a Cartesian grid.
Our fiducial resolution is $N = 256$.
We use periodic boundary conditions in all directions.
Turbulence is continuously driven using the standard Ornstein-Uhlenbeck process \citep{Schmidt2006, Federrath2010}.
We explore three different driving modes:
purely solenoidal ($b=1/3$), purely compressive ($b=1$), and a mixed mode ($b=0.5$).
The target turbulent root-mean-square (RMS) velocity is set to be $v_{\rm rms} = 1$,
which allows us to fix the turbulent crossing time $t_{\rm turb} = L / v_{\rm rms} = 1$ in all simulations.
We vary the sound speed to systematically explore seven different turbulent Mach numbers $\mathcal{M} \equiv v_{\rm rms} / c_{\rm s} = $ 2, 3, 5, 10, 20, 30, and 50.
The actual Mach number measured in the simulations is close to but not exactly the same as the target Mach number (see Table~\ref{tbl:machmeasured}).
To discard the initial transient phase, we first run a pre-simulation for one turbulent timescale (i.e., up to $t=1$) and then use the final snapshot as the initial conditions to run the actual simulation from $t=0$ to $t \sim 10^2$.

\begin{deluxetable}{cccc}\label{tbl:machmeasured}
     \centering
     \tablecaption{
     Target and actual (RMS) Mach numbers in our turbulence simulations with mixed ($b=0.5$), solenoidal ($b=1/3$), and compressive ($b=1$) driving modes.
     Actual Mach numbers are measured as the median of $v_{\rm rms}/c_s$ for $t \geq 1$ to ensure that the system has reached a steady state.
     }
    \tablehead{
     $\mathcal{M}_{\rm target}$ & $\mathcal{M}$ (mixed) & $\mathcal{M}$ (solenoidal) & $\mathcal{M}$ (compressive)
     }
     \startdata  
     2 & 2.5 & 3.0 & 1.3 \\
     3 & 3.9 & 4.1 & 2.1 \\
     5 & 5.8 & 7.1 & 3.9 \\
     10 & 11.0 & 12.6 & 8.3 \\
     20 & 21.2 & 23.6 & 15.8 \\
     30 & 30.9 & 35.2 & 23.1 \\
     50 & 49.0 & 54.6 & 38.8 \\
    \enddata
\end{deluxetable}

For each type of reaction,
we follow seven chemical species with different reaction coefficients $R_i = 10^{-3}$, $10^{-2}$, $10^{-1}$, $1$, $10^{1}$, $10^{2}$, and $10^{3}$.
We integrate the rate equations separately using the analytic solutions (Eqs.~\ref{eq:xsol_2nd}, \ref{eq:xsol_1st}, and \ref{eq:xsol_3rd}).
We set the initial abundance to $x = 1$ everywhere.
Simulation snapshots are stored from $t = 10^{-4}$ to $t=10^2$ at evenly spaced time intervals on a log scale with $\sim$100 snapshots per simulation.
The chemical species are advected with the fluid as passive scalars.
While our fiducial runs do not include sub-grid diffusion for these scalars, Appendix~\ref{app:diff} demonstrates that our results are insensitive to this choice.


\section{Results}\label{sec:results}

\subsection{Density Clumping Factor}

\begin{figure}
	\centering
	\includegraphics[width=0.99\linewidth]{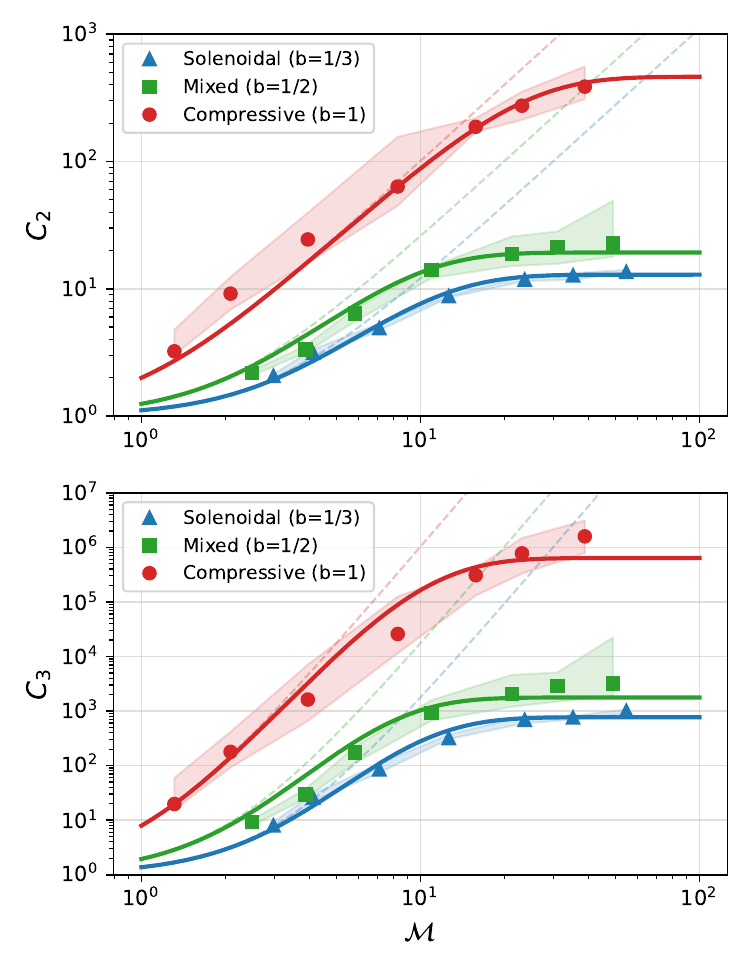}
	\caption{
        \textit{Upper panel}:
        the two-body clumping factor ($C_2 \equiv \avM{\rho}/\avV{\rho}$) as a function of the measured Mach number ($\mathcal{M}$) for solenoidal (blue triangles), mixed (green squares), and compressive (red circles) driving modes.
        The symbols indicate the median values of $C_2$ while the shaded regions bracket the 25th and 75th percentiles across snapshots.
        The solid curves show the fitting formula
        while the dashed curves show the predictions from log-normal PDF theory.
        \textit{Lower panel}:
        same as the upper panel but for the three-body clumping factor ($C_3 \equiv \avM{\rho^2}/\avV{\rho}^{\,2}$).
        Clumping factors saturate at $\mathcal{M}\gtrsim 10$.
    }
	\label{fig:cclump_driving_absolute}
\end{figure}

Before examining the mixing efficiency, we first quantify the density variance of the turbulent medium, which is completely independent of chemical reaction.
Fig.~\ref{fig:cclump_driving_absolute} shows $C_2$ (upper panel) and $C_3$ (lower panel) against the measured RMS Mach number for solenoidal (blue triangles), mixed (green squares), and compressive (red circles) driving modes.
The dashed curves show the predictions from the log-normal PDF theory: $C_2 = 1 + b^2 \mathcal{M}^2$ and $C_3 = (1+b^2\mathcal{M}^2)^3$.

At low Mach numbers,
clumping factors closely follow log-normal PDF theory, consistent with previous simulations.
However,
at high Mach numbers ($\mathcal{M} \gtrsim 10$),
clumping factors begin to saturate and fall significantly below the log-normal PDF prediction.
This deficit is more pronounced for $C_3$ than $C_2$, as the higher-order moments are more sensitive to the shape of the density PDF.
The saturation at high Mach numbers results from intermittency, as the density PDF becomes non-Gaussian in a highly supersonic turbulent medium \citep{Hopkins2013b}.
Convergence tests in Appendix~\ref{app:conv} show that the differences between $N=128$ and $N=256$ are sufficiently small that we expect the saturation to be physically robust.

To quantify the deviation from the log-normal PDF theory, we introduce an \textit{effective Mach number} that saturates at high $\mathcal{M}$:
\begin{equation}\label{eq:M_eff}
    \mathcal{M_{\rm eff}} = \mathcal{M}_0 \tanh\!\left( \frac{\mathcal{M}}{\mathcal{M}_0} \right),
\end{equation}
where $\mathcal{M}_0$ is a free parameter. 
The clumping factor thus becomes 
\begin{equation}\label{eq:cclumpfit}
    C_n = (1 + b^2 \mathcal{M}_{\rm eff}^2)^{n(n-1)/2} \quad \text{for}~ n=2,3
\end{equation}
We fit $\mathcal{M}_0$ for three driving modes individually with the least-squares method. The best-fit $\mathcal{M}_0$ are shown in Table~\ref{tab:M0fit}.
The solid curves in Fig.~\ref{fig:cclump_driving_absolute} show the fitting formula of Eq.~\ref{eq:cclumpfit},
which accurately captures the log-normal behavior at low $\mathcal{M}$ and the intermittency-driven saturation at high $\mathcal{M}$ for all three driving modes.

\begin{deluxetable}{lcc}\label{tab:M0fit}
	\tablecaption{
        Best-fit values for the saturation parameter $\mathcal{M}_0$ in Eq.~\ref{eq:cclumpfit}.
	}
	\tablewidth{0pt}
	\tablehead{
		\colhead{driving mode}   &
		\colhead{$\mathcal{M}_0(C_2)$}   &
		\colhead{$\mathcal{M}_0(C_3)$}   
	}
	\startdata  
    solenoidal  ($b=1/3$) & 10.4 & 8.6\\
    compressive ($b=1$)   & 21.5 & 9.2\\
    mixed       ($b=1/2$) &  8.7 & 6.7\\
	\enddata
\end{deluxetable}





\begin{figure*}
	\centering
	\includegraphics[width=1\linewidth]{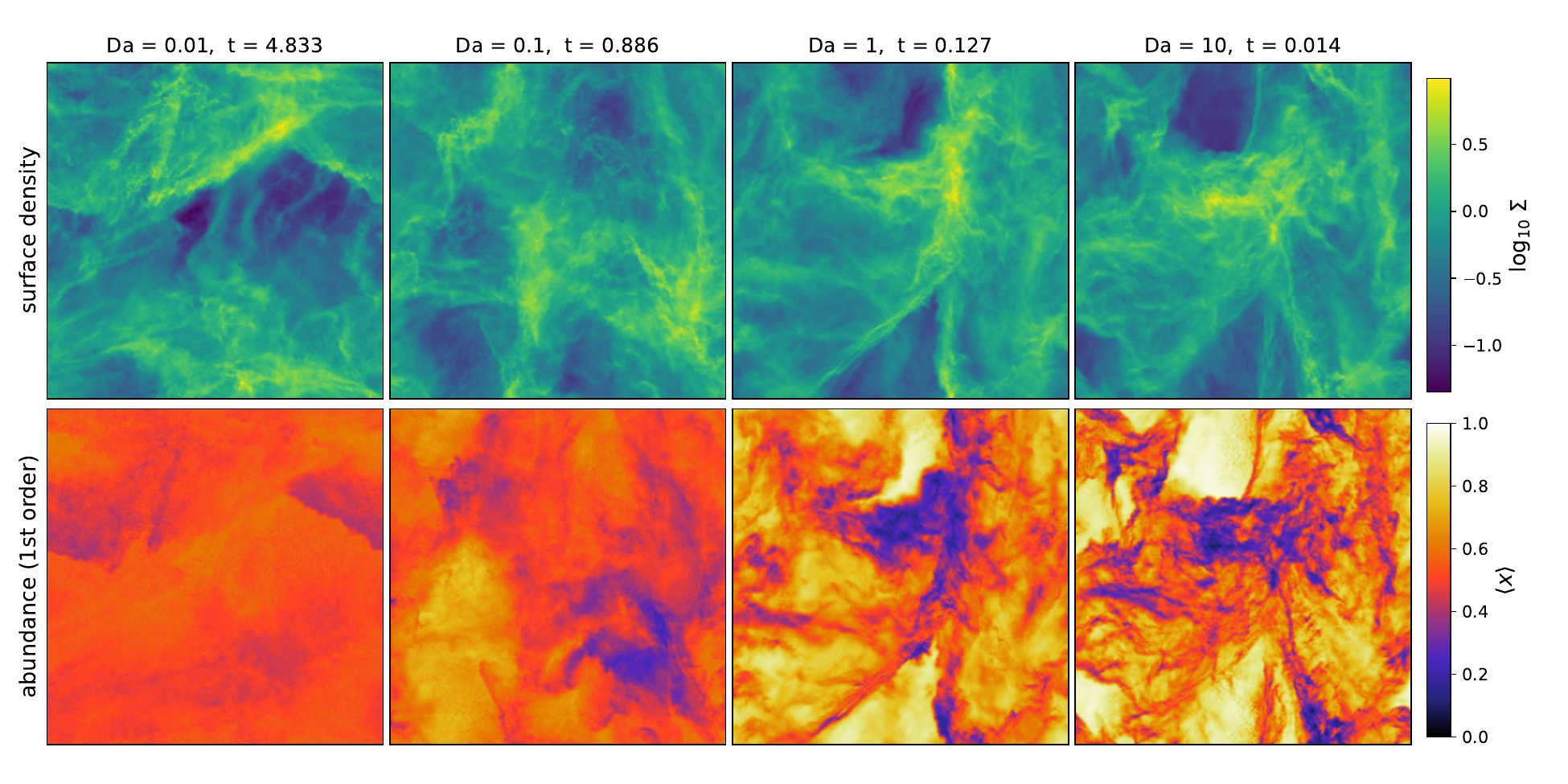}
	\caption{
        Snapshots of the turbulence simulations at the time when the mass-weighted global average chemical abundance of the 1st-order reaction reaches $\sim 0.5$.
        The upper and lower rows show the gas surface density and the local chemical abundance, respectively.
        Panels from left to right show simulations with the Damk\"ohler number ${\rm Da}\equiv t_{\rm turb} / \tau_1 = $ 0.01, 0.1, 1, and 10. 
        Abundance inhomogeneity increases with Da as chemistry outcompetes turbulent mixing.
    }
	\label{fig:surfacedensitygrid}
\end{figure*}

\subsection{Competition between Mixing and Reactions}

The abundance inhomogeneity is governed by the competition between turbulent mixing and the density-dependent chemical reactions.
This competition can be quantified by the \textit{Damk\"ohler number} of the system:
\begin{equation}
    \text{Da} \equiv \frac{t_{\rm turb}}{\tau(\avV{\rho})} = \frac{1}{R_i},
\end{equation} 
as $t_{\rm turb} = \avV{\rho} = 1$ in our setup.
When Da $\ll 1$, turbulent mixing occurs much faster than chemical reactions, and we expect the medium to be well-mixed.
In contrast, when Da $\gg 1$, reactions occur much faster than mixing, and we expect the abundance to be highly inhomogeneous.

Fig.~\ref{fig:surfacedensitygrid} shows snapshots of the turbulence simulations at the time when the mass-weighted average abundance of the 1st-order reaction reaches $\avM{x} \sim 0.5$.
The upper and lower rows show the gas surface density and the local chemical abundance, respectively.
Panels from left to right show simulations with ${\rm Da} = $ 0.01, 0.1, 1, and 10.
Since reactions occur much faster in high-Da cases, the four panels are sampled at different times at $t/t_{\rm turb} \approx 4.8, 0.89, 0.13, 0.014$ from left to right.

The surface density maps appear qualitatively similar across all runs since the system has reached a steady state from $t=0$ (thanks to the pre-simulations) and remains so throughout the simulations.
The abundance maps display a qualitative trend consistent with our intuition.
At Da = 0.01 where chemistry is slow (compared to mixing), there is sufficient time for turbulence to homogenize the abundance field.
As such, the abundance map is nearly constant everywhere.
As Da increases, the reaction becomes faster, and its density-dependent nature depletes the reactant faster in the dense gas, increasing the inhomogeneity.
At the highest Da, the reaction is so fast that the system is almost ``frozen in time'' as turbulence has no time to redistribute the heavily depleted dense gas.
This establishes a clear anti-correlation between density and abundance, as shown by comparing the upper and lower panels.

\begin{figure*}
	\centering
	\includegraphics[width=0.99\linewidth]{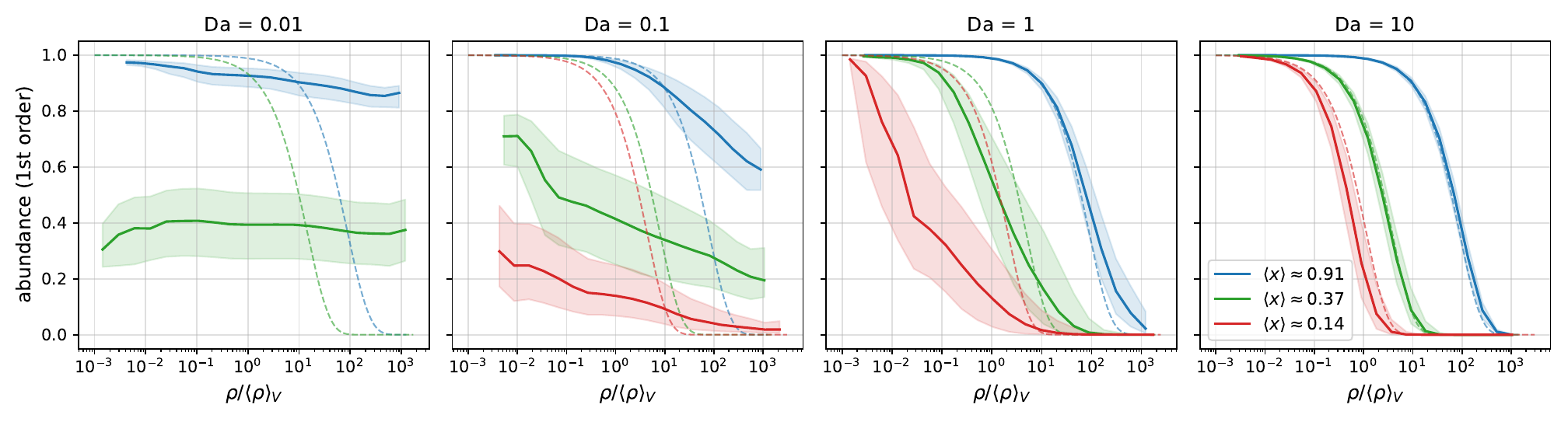}
	\caption{
        Chemical abundance of the 1st-order reaction as a function of gas density normalized to the volume-weighted density for Da $=$ 0.01, 0.1, 1, and 10 from left to right,        
        shown in blue, green, and red when the global average chemical abundance is close to 0.91, 0.37, and 0.14, respectively.
        The solid curves show the median value within a density bin while the shaded area brackets the 25th and 75th percentiles.
        The dashed curves show the analytic solutions at a fixed density.
    }
	\label{fig:abund_vs_rho_evolution_Mach10}
\end{figure*}

\begin{figure*}
	\centering
	\includegraphics[width=0.99\linewidth]{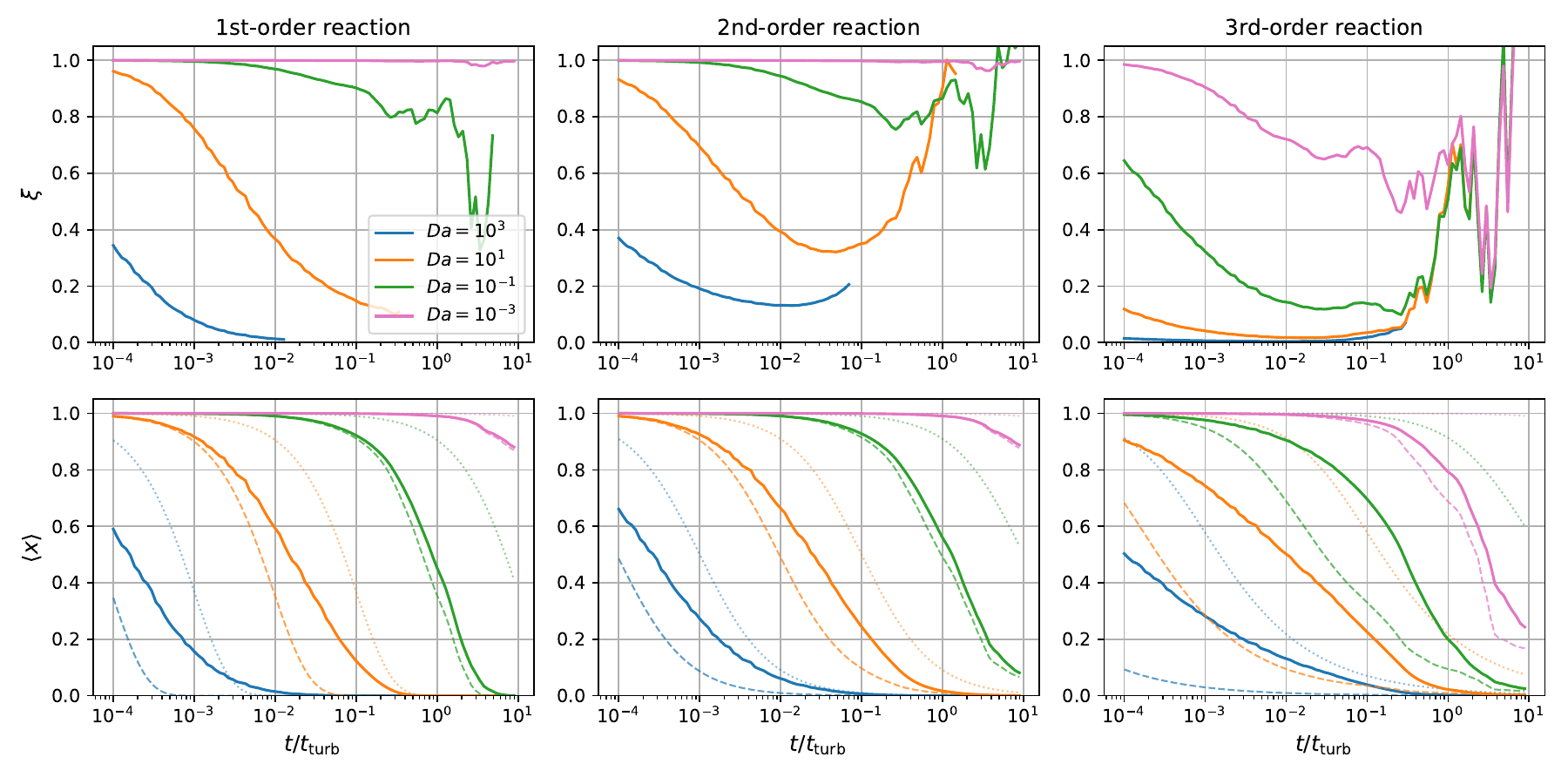}
	\caption{
        Time evolution of the mixing efficiency $\xi$ (\textit{upper panels})
        and mass-weighted average abundance $\avM{x}$ (\textit{lower panels})
        in the $\mathcal{M} = 10$ simulation
        with Da $= 10^{3}$ (blue), $10^{1}$ (orange), $10^{-1}$ (green), and $10^{-3}$ (pink)
        for the 1st-, 2nd-, and 3rd-order reaction from left to right.
        The dashed curves show the ``perfect-mixing'' limit, computed by numerically integrating the rate equation with $\xi=1$ using the measured clumping factor $C_n(t)$. 
        The dotted curves show the analytic solution in a uniform medium ($C_n=1$, $\xi=1$).
    }
	\label{fig:xi_abund_time_N256_Mach10}
\end{figure*}

To examine the density-abundance anti-correlation more quantitatively,
we show in Fig.~\ref{fig:abund_vs_rho_evolution_Mach10} the chemical abundance of the 1st-order reaction as a function of density for Da $=$ 0.01, 0.1, 1, and 10 from left to right. 
Each panel shows three different evolutionary stages sampled at $\avM{x} \approx 0.91$ (blue), 0.37 (green), and 0.14 (red), corresponding to 0.1, 1, and 2 reaction timescales, respectively.
The dashed curves show the analytic solution $x=e^{-t/\tau}$ assuming gas remains at fixed densities at all times without any turbulent transport (the ``frozen'' limit).

At low Da, the abundance is nearly flat across the entire density range, indicating that efficient mixing has redistributed the gas and erased the abundance-density anti-correlation that local chemistry would otherwise produce.
The large scatter (indicated by the shaded area) reflects the stochastic nature of mixing, as fluid elements at a given density can have very different Lagrangian histories.
As Da increases, the anti-correlation systematically increases.
At Da = 10, the reaction occurs so fast that gas is essentially frozen, and there is no time for turbulent mixing. 
Therefore,
the simulation closely follows the frozen, no-mixing analytic solution across all densities.
In addition, the scatter becomes very narrow, indicating all fluid elements at the same density have the same history (as they are frozen).


\begin{figure*}
	\centering
	\includegraphics[width=0.99\linewidth]{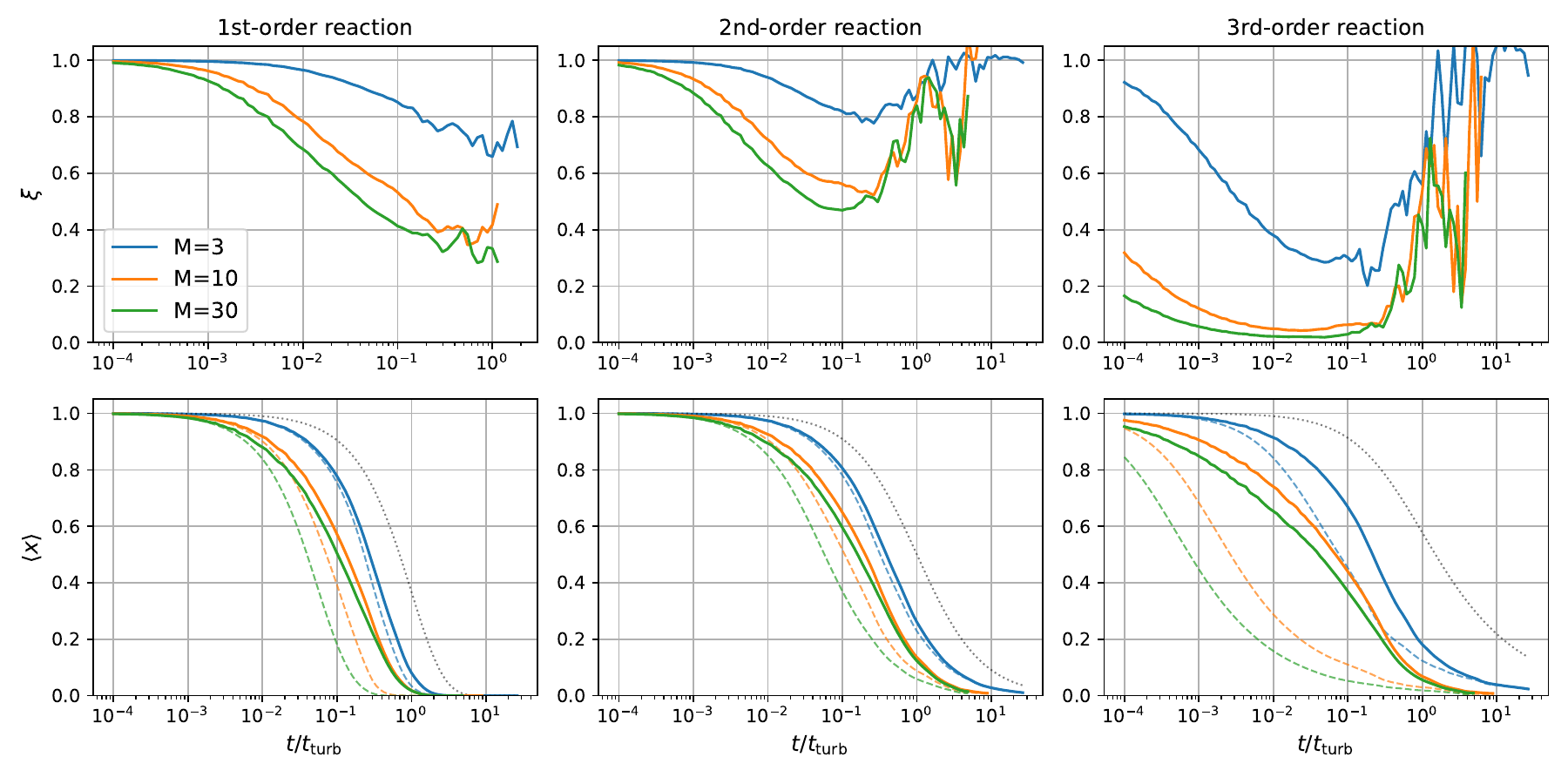}
	\caption{
        Similar to Fig.~\ref{fig:xi_abund_time_N256_Mach10} but with a fixed Da = 1 and varying Mach numbers $\mathcal{M} = 3$ (blue), 10 (orange), and 30 (green).
    }
	\label{fig:xi_abund_time_N256_Da10}
\end{figure*}

Having established the density-abundance correlation, we now examine how the mixing efficiency ($\xi$) evolves with time, as shown in the upper panels of Fig.~\ref{fig:xi_abund_time_N256_Mach10} for the $\mathcal{M} = 10$ simulation with Da $= 10^{3}$ (blue), $10^{1}$ (orange), $10^{-1}$ (green), and $10^{-3}$ (pink) for all three reaction types from left to right.
The evolution after $\avM{x}$ drops below 0.01 is discarded as the reaction is essentially complete globally.
At early times, $\xi \to 1$ for all reactions even if the reactions are density-dependent as long as the reactant is still abundant such that $x \approx 1$ everywhere.
Once the reactant is sufficiently depleted, which occurs in the dense gas first, the local reaction slows down unless it can be replenished by turbulence.
Therefore, species with Da $\gg 1$ (fast chemistry, inefficient mixing) quickly develop the density-abundance anti-correlation and so $\xi$ drops accordingly.
In contrast, species with Da $\ll 1$ (slow chemistry, efficient mixing) are depleted slowly, leaving sufficient time for turbulence to redistribute the abundance field and limit the decline in $\xi$.

\begin{figure*}
	\centering
	\includegraphics[width=0.99\linewidth]{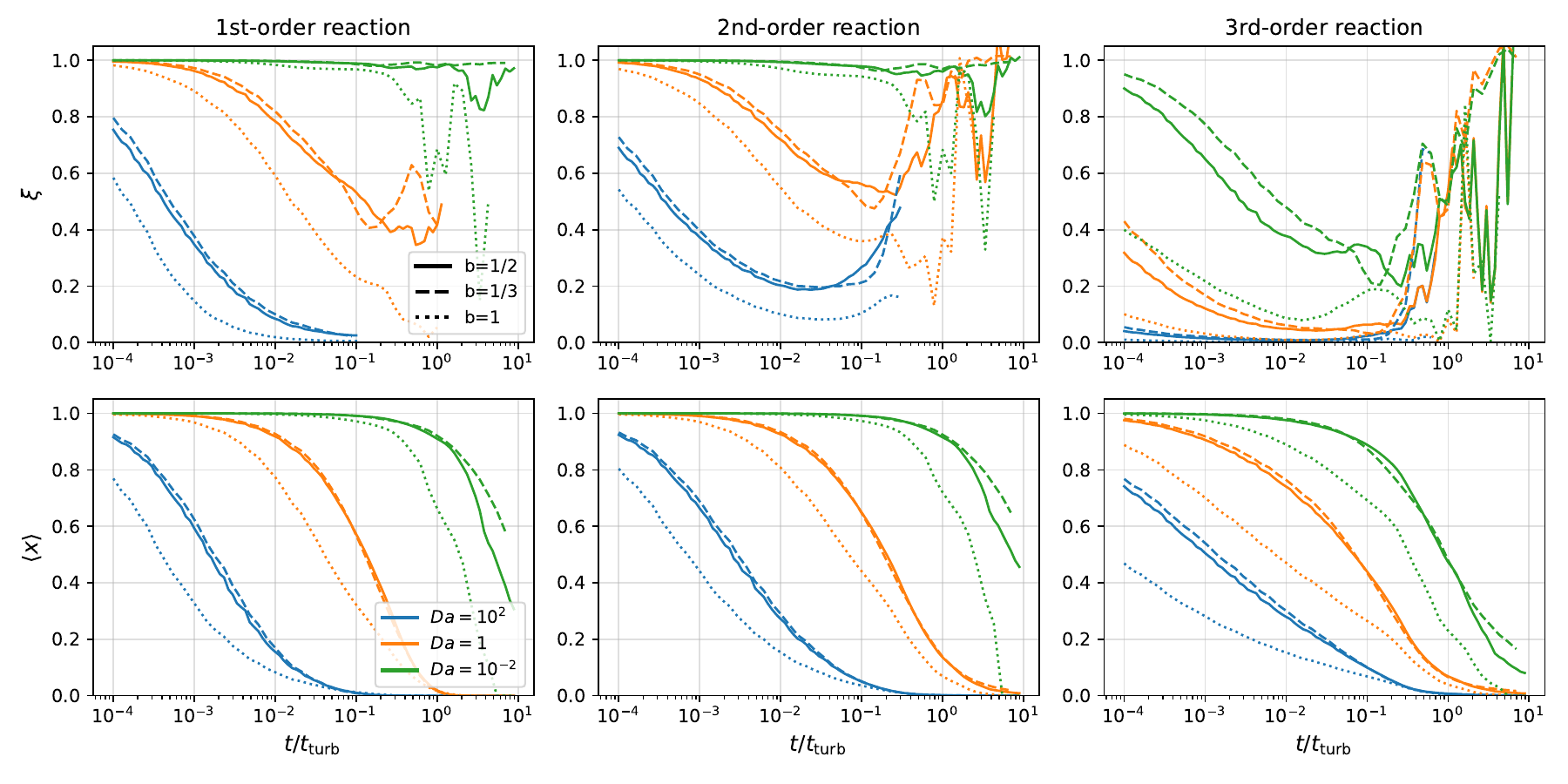}
	\caption{
        Similar to Fig.~\ref{fig:xi_abund_time_N256_Mach10}
        with $\mathcal{M} = 10$ and Da $= 10^2$ (blue), $1$ (orange), and $10^{-2}$ (green) 
        but with varying forcing parameters $b = 1/3$ (dashed), $1/2$ (solid), and $1$ (dotted).
    }
	\label{fig:xi_abund_driving_all_Mach10}
\end{figure*}

The corresponding evolution of the global abundance is shown in the lower panels of Fig.~\ref{fig:xi_abund_time_N256_Mach10} 
For reference,
the dashed curves show the ``perfect-mixing'' limit, obtained by numerically integrating the rate equation with $\xi=1$ using the measured clumping factor $C_n(t)$. 
This is to ensure that the reaction boost is correctly captured as $C_n$ fluctuates with time even if the system is in a steady state.
The dotted curves show the analytic solution in a uniform medium ($C_n=1$, $\xi=1$).

The difference between the dashed and dotted curves reflects the reaction boost in the perfect-mixing limit (as commonly adopted by cosmological simulations).
However, this reaction boost is \textit{overestimated} due to incomplete mixing, whose cumulative effect is quantified by the difference between the solid and dashed curves.
The effect is larger at high Da, indicating a consistent picture as in $\xi$, as fast chemistry depletes reactants in dense gas before turbulence can homogenize the abundance, suppressing the reaction boost.
Interestingly, at Da $= 10^3$ for the 1st-order reaction, this suppression is so strong at late times that $\avM{x}$ declines even more slowly than the uniform-medium limit (dotted curves), which occurs as all reactants at densities larger than $\avV{\rho}$ are essentially depleted.

The evolution is qualitatively similar across the three reaction types: higher Da leads to lower $\xi$ and thus a stronger reaction suppression.
The suppression is significantly stronger for the 3rd-order reaction due to its $\rho^2$-dependent rate equation, making it more sensitive to density variation and thus generating a stronger density-abundance anti-correlation.
In addition, for both 2nd- and 3rd-order reactions, $\xi$ does not monotonically decline with time. 
Instead, it first declines to a minimum and then increases again at late times.
This is because $\xi$ actually contains two effects for higher-order reactions.
Take the 2nd-order reaction as an example (the 3rd-order reaction is qualitatively similar), we can express it as
\begin{equation}\label{eq:Ushape}
    \xi_2 \equiv \frac{\avM{\rho x^2}}{\avM{\rho}\avM{x^2}} \frac{\avM{x^2}}{\avM{x}^{2}}.
\end{equation}
The first term captures the density-abundance anti-correlation discussed above, which declines monotonically with time.
On the other hand, the second term describes the spatial variance of $x$, which is small at early times when $x$ is uniform and gradually increases with time as reactants are depleted in dense gas.
This is because the 2nd-order reaction rate locally scales with $x^2$ as it depends on one reactant atom to find another one, both of which are depleted at the same rate (as opposed to the 1st-order reaction where there is an infinite reservoir of the collision partner).
When the reactant is completely depleted in dense gas, the residual reactant will be concentrated in the diffuse gas, leading to more efficient reaction than if $x$ is homogenized by turbulence.
The combined effect of these two factors leads to the U-shaped trend in the $\xi$-evolution.

\subsection{Mach Number Dependence}

Fig.~\ref{fig:xi_abund_time_N256_Da10} shows similar diagnostic as Fig.~\ref{fig:xi_abund_time_N256_Mach10} but with a fixed Da = 1 for three Mach numbers $\mathcal{M} = 3$, 10, and 30.
At a fixed reaction rate, the medium is more compressible at higher $\mathcal{M}$, leading to a broader density distribution.
This means that the density-dependent reaction has a higher contrast, creating the density-abundance anti-correlation and yielding a smaller $\xi$.
Therefore, there is a significant difference in $\xi$ between $\mathcal{M}=3$ and $\mathcal{M}=10$.
In the subsonic regime ($\mathcal{M}< 1$),
the density variations generated by turbulence are very weak and the medium remains nearly uniform, hence $\xi\to 1$ by construction even with fast chemistry.
The $\mathcal{M}=10$ and $\mathcal{M}=30$ cases are much more similar, which is a direct consequence of saturation at high Mach numbers.
Comparing across the three reaction types,
the 3rd-order reaction shows the most significant $\mathcal{M}$-dependence as its rate equation depends on $\rho^2$, making it more sensitive to the density structure.

\begin{figure*}
	\centering
	\includegraphics[width=0.99\linewidth]{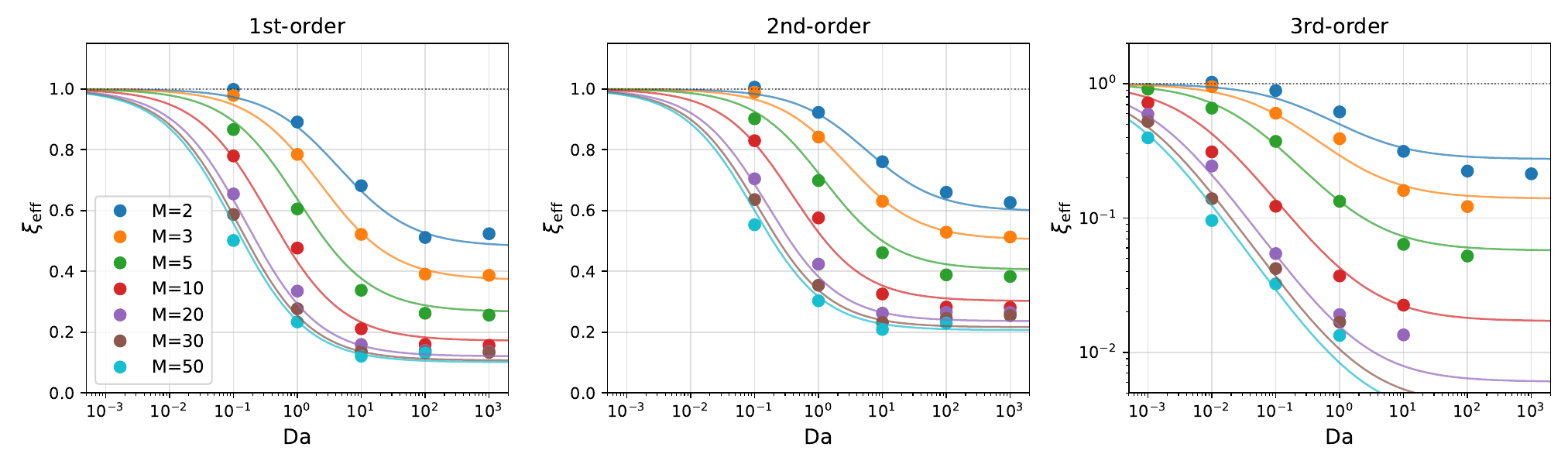}
	\caption{
        Effective mixing efficiency $\xi_{\rm eff}$ as a function of Da with varying $\mathcal{M} = $ 2, 3, 5, 10, 20, 30, and 50,
        overplotted with the fitting formula (Eq.~\ref{eq:xi_eff}) in solid curves.
        The fitting formula can be implemented as a sub-grid model in cosmological simulations.
    }
	\label{fig:xi_eff_fit_vs_Da_log}
\end{figure*}

\subsection{Driving Mode Dependence}

So far we have focused on the mixed driving mode.
Here we investigate the effect of driving mode by exploring two extreme cases: purely compressive ($b=1$) and purely solenoidal ($b=1/3$) driving.
Fig.~\ref{fig:xi_abund_driving_all_Mach10} shows similar diagnostic as Fig.~\ref{fig:xi_abund_time_N256_Mach10} for $\mathcal{M} = 10$ and Da $= 10^2$ (blue), $1$ (orange), and $10^{-2}$ (green) but with $b = 1/3$ (dashed), $1/2$ (solid), and $1$ (dotted).

The driving mode affects the width of the density PDF, $\sigma_s^2 = \ln{(1 + b^2 \mathcal{M}^2)}$ \citep{Federrath2008,Federrath2010}.
At a fixed $\mathcal{M}$, compressive driving produces a significantly broader density PDF than solenoidal driving.
This has two competing effects.
On one hand, compressive driving makes the medium clumpier with denser gas to boost reactions (higher $C_n$).
On the other hand, the larger density contrast it creates leads to more efficient local depletion of reactants in dense gas, strengthening the density-abundance anti-correlation and thus lowering $\xi$.
The former dominates in all cases, and the net effect is that compressive driving leads to a faster global reaction than solenoidal and mixed driving.
The mixed-mode driving falls in between the two extreme modes,
but it behaves much more similarly to the solenoidal mode than the compressive mode.
This is consistent with the trend in Fig.~\ref{fig:cclump_driving_absolute}, where mixed-mode driving leans more toward the solenoidal driving.

\subsection{Effective Mixing Efficiency}


Figs.~\ref{fig:xi_abund_time_N256_Mach10} and ~\ref{fig:xi_abund_time_N256_Da10} establish that $\xi$ depends on both Da and $\mathcal{M}$.
However, $\xi$ also depends on time in a nontrivial way that is difficult to parametrize.
Therefore, we define the \textit{effective mixing efficiency} ($\xi_{\rm eff}$) as a single parameter to describe the cumulative effect of incomplete mixing. 
For simplicity, we focus on mixed-driving simulations for parametrization.
For each simulation,
we measure the ``crossing-time'' $t_{\rm cr}$ at which $\avM{x}$ drops below a critical value.
This value is chosen as the abundance at two chemical times in the analytic solution, i.e., 
$x_{\rm cr} = e^{-2}$ (1st order), 1/3 (2nd order), and $1/\sqrt{5}$ (3rd order).
We contrast this crossing time to the crossing time where the abundance is numerically integrated using the actual clumping factor assuming perfect mixing (see the dashed curves in Fig.~\ref{fig:xi_abund_time_N256_Mach10}) and define
\begin{equation}
    \xi_{\rm eff} \equiv \frac{t_{\rm cr}(\xi=1)}{t_{\rm cr}}.
\end{equation}
This parameter acts as a cumulative proxy for reaction suppression.

Fig.~\ref{fig:xi_eff_fit_vs_Da_log} shows the measured $\xi_{\rm eff}$ as a function of Da for all seven Mach numbers (colored circles) and the three reaction types.
The 3rd-order reaction is shown on a log scale on the $y$-axis to accommodate the wider dynamic range.
Note that while the legend shows the target Mach numbers for readability,
the fit uses the measured Mach numbers. 
The results show asymptotic behaviors on both extreme ends of Da.
For Da $\ll 1$, chemistry is slow and mixing dominates, and thus $\xi_{\rm eff} = 1$ (perfectly mixed).
For Da $\gg 1$, chemistry is so fast that the gas is essentially frozen, and thus $\xi_{\rm eff}$ saturates to a fixed value $\xi_{\infty}(\mathcal{M})$.
This value is $\mathcal{M}$-dependent, as higher $\mathcal{M}$ creates a larger range of densities.
At intermediate Da,
$\xi_{\rm eff}$ gradually transitions between the two extremes.
The transition occurs near a critical value Da$_0$ that decreases with increasing $\mathcal{M}$.
This is again due to the broader density PDF at higher $\mathcal{M}$, which makes it easier for chemistry to dominate over mixing and develop the density-abundance anti-correlation.

For practical use in cosmological simulations as a sub-grid model,
we provide a fitting formula:
\begin{equation}\label{eq:xi_eff}
\xi_{\rm eff}({\rm Da}, \mathcal{M}) = \xi_\infty + \frac{1 - \xi_\infty}{1 + ({\rm Da} / {\rm Da_0})^f}
\end{equation}
where 
\begin{equation}
\xi_\infty  = \frac{a}{1 + \beta \mathcal{M}^{\prime c}_{\rm eff}}
\end{equation}
is the high-Da floor,
Da$_0 = d \mathcal{M}^{\prime e}_{\rm eff}$ is the critical value of Da,
and 
$\mathcal{M}^\prime_{\rm eff}  = \mathcal{M}^\prime_0 \tanh(\mathcal{M} / \mathcal{M}^\prime_0)$
is the effective Mach number accounting for saturation.
The best-fit parameters are given in Table~\ref{tab:xi_eff_params}.
We note that the effective Mach number here (denoted with a prime) differs from that in Eq.~\ref{eq:M_eff} for clumping factors.
Empirically, $\xi_{\rm eff}$ saturates at higher $\mathcal{M}$, and using the same parameters leads to a significantly worse fit, as we found in earlier attempts.
The formula is intended to be used in the range of $\mathcal{M}\geq 1$,
where the gas is supersonically turbulent.
When extrapolating outside the fitted range $\mathcal{M}=$ 2 -- 50,
$\xi_\infty$ can slightly exceed unity at $\mathcal{M} \approx 1$
and should be capped at 1.

\begin{deluxetable}{lccccccc}\label{tab:xi_eff_params}
\tablecaption{Best-fit parameters of the fitting formula for the effective mixing efficiency $\xi_{\rm eff}$ (Eq.~\ref{eq:xi_eff}) for each reaction type from the mixed-driving simulations ($b = 0.5$).}
\tablewidth{0pt}
\centering
\tablehead{
order & $a$ & $\beta$ & $c$ & $d$ & $e$ & $f$ & $\mathcal{M}_0^\prime$
}
\startdata  
1${\rm st}$ & 3.484 & 2.913 & 0.832 & 24.78 & $-1.862$ & 0.762 & 19.1  \\
2${\rm nd}$ & 3.062 & 2.423 & 0.581 & 33.24 & $-1.944$ & 0.783 & 20.4  \\
3${\rm rd}$ & 1.996 & 0.909 & 2.105 &  4.46 & $-2.860$ & 0.712 & 22.9  \\
\enddata
\end{deluxetable}

\section{Discussion}
\label{sec:discussion}


\subsection{Comparison with Previous Work}

It is well established that the density PDF in an isothermal, supersonically turbulent medium follows a log-normal distribution \citep{Vazquez-Semadeni1994, Padoan1997, Passot1998, Federrath2008, Federrath2010},
and our clumping factors at low Mach numbers are fully consistent with this theoretical framework.
In addition,
the deviation from the log-normal PDF at high Mach numbers has also been shown, and fitting formulae for the density distribution have been provided \citep{Konstandin2012, Hopkins2013b, Nolan2015, Mocz2019, Scannapieco2026}.
However, none of these provided a closed-form fitting formula for the clumping factors, unlike our effective Mach number prescription (Eq.~\ref{eq:M_eff}) that is designed for practical use in modeling sub-grid chemical reactions.


The most relevant work addressing the effect of incomplete mixing on suppressing chemical reactions is \citetalias{Micic2012}.
The key difference is that their simulations included self-consistent cooling, heating, and various destruction mechanisms.
Despite being more physically sophisticated,
their system is not scale-free.
As such,
their results are limited to the specific initial conditions explored, which cannot be generalized to cosmological simulations.
The key contribution of our work is identifying the scale-free nature of the problem under isothermal, well-shielded conditions,
reducing the parameter space to just two dimensions: $\mathcal{M}$ and Da.
This crucial step allows us to scan the parameter space efficiently, leading to our simple fitting formula that can be implemented in cosmological simulations.
We note that while ignoring destruction mechanisms such as radiation shielding is a reasonable approximation at high densities where the bulk of the molecular gas resides,
it breaks down in the diffuse gas where formation and destruction are competing with each other.
We plan to investigate the effect of radiation shielding in future work.

\subsection{Clumping Factor as a Sub-grid Model}

The use of clumping factors to boost unresolved chemical reactions is widespread in cosmological and galaxy-formation simulations, mostly focusing on H$_2$ formation.
Early applications were mostly done by assuming a constant clumping factor throughout the simulations \citep{Gnedin2009, Kuhlen2012, Christensen2012, Pallottini2017}.
More recently,
\citet{Lupi2018} adopted an $\mathcal{M}$-dependent clumping factor to boost H$_2$ formation, estimated locally from the velocity dispersion of gas.
\citet{Hopkins2023} further extended the model for both two-body and three-body reactions in the recent cosmological simulations FIRE-3.
These models implicitly assume instantaneous mixing of chemical abundances and could significantly overestimate the reaction rates.
We offer an easy-to-implement fix by replacing the clumping factor with our empirical $\xi_{\rm eff} C_n$ (Eqs.~\ref{eq:M_eff}, \ref{eq:cclumpfit} and \ref{eq:xi_eff}),
where both $\mathcal{M}$ and Da can be estimated locally from the velocity dispersion.

Molecular clouds in the Milky Way and similar star-forming galaxies follow Larson's relation \citep{Larson1981, Heyer2004}:
$\sigma_{\rm v} \sim {\rm 1\ km\ s^{-1}} \Delta^{0.5}$,
where $\sigma_{\rm v}$ is the velocity dispersion, 
$\Delta$ is the characteristic cloud size in units of pc.
The resulting turbulent timescale is thus $t_{\rm turb} \sim {\rm 1\ Myr} \Delta^{0.5}$.
In comparison,
the H$_2$ formation time is $\sim$ 1~Gyr~$n^{-1}$ at solar metallicity, 
where $n$ is the hydrogen number density in units of cm$^{-3}$.
The typical Damk\"ohler number in molecular clouds is therefore ${\rm Da} \sim 10^{-3} n \Delta^{0.5}$.
For cosmological simulations with a spatial resolution of $\Delta=100$~pc,  
this corresponds to Da $\sim$ 1 -- 10 at typical densities of H$_2$ formation, $n = 10^2$ -- $10^3 ~{\rm cm^{-3}}$.
This lies right in the transition regime where the mixing efficiency is neither unity nor fully saturated to $\xi_{\infty}$, and the effect ranges from only a mild correction of 10\% to a drastic ten-fold suppression, depending on the exact values of $\mathcal{M}$ and Da.
For Lagrangian codes, the spatial resolution is adaptive:
$\Delta\sim 6~{\rm pc}~(m_{\rm g}/n)^{1/3}$ for a kernel radius that contains by 32 neighbors, where $m_{\rm g}$ is the particle mass in units of M$_\odot$.
This leads to $t_{\rm turb}\sim 2.4~{\rm Myr} (m_{\rm g}/n)^{1/6}$
and thus Da~$\sim 2.4\times 10^{-3} n^{5/6} m_{\rm g}^{1/6}$,
rather insensitive to the mass resolution.
For $m_{\rm g}\sim 10^3~{\rm M_\odot}$,
Da $\sim$ 0.35 -- 2.4 for $n = 10^2$ -- $10^3 ~{\rm cm^{-3}}$, 
also in the transition regime.


\section{Summary}
\label{sec:summary}

We have conducted hydrodynamical simulations of isothermal, supersonically driven turbulence to quantify how turbulence enhances chemical reactions.
We demonstrate that the global reaction rate in a turbulent medium is fully characterized by two dimensionless parameters:
the Mach number ($\mathcal{M}$), which determines the density distribution, and the Damk\"ohler number (Da), which quantifies the competition between turbulent mixing and the local, density-dependent reaction rate.
Our simulations span $\mathcal{M} = 2$ -- 50,
Da = $10^{-3}$ -- $10^{3}$,
three choices of forcing parameter ($b$ = 1/3, 0.5, and 1),
and three reaction types (first-, second-, and third-order).


The density-abundance anti-correlation is driven by the competition between density-dependent reactions that preferentially deplete reactants in dense gas and turbulent mixing that homogenizes the abundance field (Fig.~\ref{fig:abund_vs_rho_evolution_Mach10}).
This anti-correlation (quantified by Da) and the density distribution of gas (determined by $\mathcal{M}$) control the mixing efficiency $\xi$.
At Da $\ll 1$, turbulent mixing completely smooths out the abundance field and so $\xi \approx 1$.
At Da $\gg 1$, reactions are so rapid that the gas is essentially frozen throughout the entire reaction process, and hence $\xi \ll 1$ (Fig.~\ref{fig:xi_abund_time_N256_Mach10}).
On the other hand, the $\mathcal{M}$-dependence simply reflects the underlying density structure.
Higher $\mathcal{M}$ broadens the density PDF, leaving more room for the density-abundance anti-correlation to develop and thus leading to smaller $\xi$ (Fig.~\ref{fig:xi_abund_time_N256_Da10}).
For the same reason, $\xi$ also depends on the turbulent forcing mode. 
At fixed $\mathcal{M}$, compressive driving produces a lower $\xi$ than solenoidal driving due to its broader density PDF (Fig.~\ref{fig:xi_abund_driving_all_Mach10}).

Mixing efficiency evolves over time as the reaction proceeds.
At early times, reactants are abundant at all densities and so $\xi \approx 1$.
As time evolves, $\xi$ gradually decreases because reactants are preferentially depleted in dense gas.
For 1st-order reactions, $\xi$ decreases monotonically.
For 2nd- and 3rd-order reactions,
$\xi$ increases again at late times due to spatial variance of the abundance field (see Eq.~\ref{eq:Ushape}),
leading to a U-shaped evolution.


For practical use as a sub-grid model,
we quantify the cumulative effect of incomplete mixing with an \textit{effective} mixing efficiency $\xi_{\rm eff}$, using the time at which the average abundance drops below a threshold value as a proxy of reaction completeness (Fig.~\ref{fig:xi_eff_fit_vs_Da_log}),
and we provide a fitting formula for $\xi_{\rm eff}$ as a function of $\mathcal{M}$ and Da (Eq.~\ref{eq:xi_eff}).
In addition, we also parametrize the clumping factors ($C_n$), which require correction at high Mach numbers ($\mathcal{M} \gtrsim 10$) due to intermittency (Fig.~\ref{fig:cclump_driving_absolute}), into fitting formulae Eqs.~\ref{eq:M_eff} and \ref{eq:cclumpfit}.
In cosmological simulations,
the sub-grid reaction rates should be enhanced by a factor of $\xi_{\rm eff} C_n$
(rather than just $C_n$ as in the traditional approach) to account for incomplete mixing.


\section*{Acknowledgments}
C.Y.H. acknowledges support from the NSTC grant 113-2112-M-002-041-MY3 and 
the National Center for High-performance Computing (NCHC) of the National Institutes of Applied Research (NIAR) in Taiwan for providing computational and storage resources.
We use Claude Code for plotting and finding the fitting formula for $\xi_{\rm eff}$.

\bibliography{literatur}{}

@article{Schmidt2006,
  author = {Schmidt, Wolfram and Hillebrandt, Wolfgang and Niemeyer, Jens C.},
  title = {Numerical dissipation and the bottleneck effect in simulations of compressible isotropic turbulence},
  journal = {Computers \& Fluids},
  volume = {35},
  number = {4},
  pages = {353--371},
  month = {May},
  year = {2006},
  doi = {10.1016/j.compfluid.2005.03.002}
}

@ARTICLE{Hu2020b,
       author = {{Hu}, Chia-Yu and {Chiang}, Chi-Ting},
        title = "{A Priori Validation of Subgrid-scale Models for Astrophysical Turbulence}",
      journal = {\apj},
         year = 2020,
        month = sep,
       volume = {900},
       number = {1},
          eid = {29},
        pages = {29},
          doi = {10.3847/1538-4357/aba2d5},
archivePrefix = {arXiv},
       eprint = {2003.13780},
 primaryClass = {astro-ph.GA},
       adsurl = {https://ui.adsabs.harvard.edu/abs/2020ApJ...900...29H}
}

@ARTICLE{Tan2021,
       author = {{Tan}, Brent and {Oh}, S. Peng and {Gronke}, Max},
        title = "{Radiative mixing layers: insights from turbulent combustion}",
      journal = {\mnras},
         year = 2021,
        month = apr,
       volume = {502},
       number = {3},
        pages = {3179-3199},
          doi = {10.1093/mnras/stab053},
archivePrefix = {arXiv},
       eprint = {2008.12302},
 primaryClass = {astro-ph.GA},
       adsurl = {https://ui.adsabs.harvard.edu/abs/2021MNRAS.502.3179T}
}

@ARTICLE{Girma2024,
       author = {{Girma}, Eden and {Teyssier}, Romain},
        title = "{A new star formation recipe for magnetohydrodynamics simulations of galaxy formation}",
      journal = {\mnras},
         year = 2024,
        month = jan,
       volume = {527},
       number = {3},
        pages = {6779-6794},
          doi = {10.1093/mnras/stad3640},
archivePrefix = {arXiv},
       eprint = {2311.10826},
 primaryClass = {astro-ph.GA},
       adsurl = {https://ui.adsabs.harvard.edu/abs/2024MNRAS.527.6779G}
}

@ARTICLE{Kretschmer2020,
       author = {{Kretschmer}, Michael and {Teyssier}, Romain},
        title = "{Forming early-type galaxies without AGN feedback: a combination of merger-driven outflows and inefficient star formation}",
      journal = {\mnras},
         year = 2020,
        month = feb,
       volume = {492},
       number = {1},
        pages = {1385-1398},
          doi = {10.1093/mnras/stz3495},
archivePrefix = {arXiv},
       eprint = {1906.11836},
 primaryClass = {astro-ph.GA},
       adsurl = {https://ui.adsabs.harvard.edu/abs/2020MNRAS.492.1385K}
}

@ARTICLE{Semenov2016,
       author = {{Semenov}, Vadim A. and {Kravtsov}, Andrey V. and {Gnedin}, Nickolay Y.},
        title = "{Nonuniversal Star Formation Efficiency in Turbulent ISM}",
      journal = {\apj},
       volume = {826},
         year = 2016,
        month = aug,
       number = {2},
          eid = {200},
        pages = {200},
          doi = {10.3847/0004-637X/826/2/200},
archivePrefix = {arXiv},
       eprint = {1512.03101},
 primaryClass = {astro-ph.GA},
       adsurl = {https://ui.adsabs.harvard.edu/abs/2016ApJ...826..200S}
}

@ARTICLE{Hennebelle2011,
       author = {{Hennebelle}, Patrick and {Chabrier}, Gilles},
        title = "{Analytical Star Formation Rate from Gravoturbulent Fragmentation}",
      journal = {\apjl},
         year = 2011,
        month = dec,
       volume = {743},
       number = {2},
          eid = {L29},
        pages = {L29},
          doi = {10.1088/2041-8205/743/2/L29},
archivePrefix = {arXiv},
       eprint = {1110.0033},
 primaryClass = {astro-ph.GA},
       adsurl = {https://ui.adsabs.harvard.edu/abs/2011ApJ...743L..29H}
}

@ARTICLE{Padoan2011,
       author = {{Padoan}, Paolo and {Nordlund}, {\r{A}}ke},
        title = "{The Star Formation Rate of Supersonic Magnetohydrodynamic Turbulence}",
      journal = {\apj},
         year = 2011,
        month = mar,
       volume = {730},
       number = {1},
          eid = {40},
        pages = {40},
          doi = {10.1088/0004-637X/730/1/40},
archivePrefix = {arXiv},
       eprint = {0907.0248},
 primaryClass = {astro-ph.GA},
       adsurl = {https://ui.adsabs.harvard.edu/abs/2011ApJ...730...40P}
}

@ARTICLE{Larson2005,
       author = {{Larson}, Richard B.},
        title = "{Thermal physics, cloud geometry and the stellar initial mass function}",
      journal = {\mnras},
         year = 2005,
        month = may,
       volume = {359},
       number = {1},
        pages = {211-222},
          doi = {10.1111/j.1365-2966.2005.08881.x},
archivePrefix = {arXiv},
       eprint = {astro-ph/0412357},
 primaryClass = {astro-ph},
       adsurl = {https://ui.adsabs.harvard.edu/abs/2005MNRAS.359..211L}
}

@ARTICLE{Klessen2016,
       author = {{Klessen}, Ralf S. and {Glover}, Simon C.~O.},
        title = "{Physical Processes in the Interstellar Medium}",
      journal = {Saas-Fee Advanced Course},
         year = 2016,
        month = jan,
       volume = {43},
        pages = {85},
          doi = {10.1007/978-3-662-47890-5_2},
archivePrefix = {arXiv},
       eprint = {1412.5182},
 primaryClass = {astro-ph.GA},
       adsurl = {https://ui.adsabs.harvard.edu/abs/2016SAAS...43...85K}
}

@ARTICLE{Dutta2015,
       author = {{Dutta}, Jayanta and {Nath}, Biman B. and {Clark}, Paul C. and {Klessen}, Ralf S.},
        title = "{The role of three-body H$_{2}$ formation in the fragmentation of primordial gas}",
      journal = {\mnras},
         year = 2015,
        month = jun,
       volume = {450},
       number = {1},
        pages = {202-208},
          doi = {10.1093/mnras/stv664},
archivePrefix = {arXiv},
       eprint = {1503.08932},
 primaryClass = {astro-ph.GA},
       adsurl = {https://ui.adsabs.harvard.edu/abs/2015MNRAS.450..202D}
}

@ARTICLE{Bovino2014,
       author = {{Bovino}, S. and {Schleicher}, D.~R.~G. and {Grassi}, T.},
        title = "{Primordial star formation: relative impact of H$_{2}$ three-body rates and initial conditions}",
      journal = {\aap},
         year = 2014,
        month = jan,
       volume = {561},
          eid = {A13},
        pages = {A13},
          doi = {10.1051/0004-6361/201322387},
archivePrefix = {arXiv},
       eprint = {1307.7567},
 primaryClass = {astro-ph.CO},
       adsurl = {https://ui.adsabs.harvard.edu/abs/2014A&A...561A..13B}
}

@ARTICLE{Abel2002,
       author = {{Abel}, Tom and {Bryan}, Greg L. and {Norman}, Michael L.},
        title = "{The Formation of the First Star in the Universe}",
      journal = {Science},
         year = 2002,
        month = jan,
       volume = {295},
       number = {5552},
        pages = {93-98},
          doi = {10.1126/science.1063991},
archivePrefix = {arXiv},
       eprint = {astro-ph/0112088},
 primaryClass = {astro-ph},
       adsurl = {https://ui.adsabs.harvard.edu/abs/2002Sci...295...93A}
}

@ARTICLE{Heyer2015,
       author = {{Heyer}, Mark and {Dame}, T.~M.},
        title = "{Molecular Clouds in the Milky Way}",
      journal = {\araa},
         year = 2015,
        month = aug,
       volume = {53},
        pages = {583-629},
          doi = {10.1146/annurev-astro-082214-122324},
       adsurl = {https://ui.adsabs.harvard.edu/abs/2015ARA&A..53..583H}
}

@ARTICLE{Heyer2004,
       author = {{Heyer}, Mark H. and {Brunt}, Christopher M.},
        title = "{The Universality of Turbulence in Galactic Molecular Clouds}",
      journal = {\apjl},
         year = 2004,
        month = nov,
       volume = {615},
       number = {1},
        pages = {L45-L48},
          doi = {10.1086/425978},
archivePrefix = {arXiv},
       eprint = {astro-ph/0409420},
 primaryClass = {astro-ph},
       adsurl = {https://ui.adsabs.harvard.edu/abs/2004ApJ...615L..45H}
}

@ARTICLE{RomanDuval2011,
       author = {{Roman-Duval}, Julia and {Federrath}, Christoph and {Brunt}, Christopher and {Heyer}, Mark and {Jackson}, James and {Klessen}, Ralf S.},
        title = "{The Turbulence Spectrum of Molecular Clouds in the Galactic Ring Survey: A Density-dependent Principal Component Analysis Calibration}",
      journal = {\apj},
         year = 2011,
        month = oct,
       volume = {740},
       number = {2},
          eid = {120},
        pages = {120},
          doi = {10.1088/0004-637X/740/2/120},
archivePrefix = {arXiv},
       eprint = {1107.4569},
 primaryClass = {astro-ph.GA},
       adsurl = {https://ui.adsabs.harvard.edu/abs/2011ApJ...740..120R}
}

@ARTICLE{Zuckerman1974,
       author = {{Zuckerman}, B. and {Evans}, II, N.~J.},
        title = "{Models of Massive Molecular Clouds}",
      journal = {\apjl},
         year = 1974,
        month = sep,
       volume = {192},
        pages = {L149},
          doi = {10.1086/181613},
       adsurl = {https://ui.adsabs.harvard.edu/abs/1974ApJ...192L.149Z}
}

@ARTICLE{Pallottini2017,
       author = {{Pallottini}, A. and {Ferrara}, A. and {Bovino}, S. and {Vallini}, L. and {Gallerani}, S. and {Maiolino}, R. and {Salvadori}, S.},
        title = "{The impact of chemistry on the structure of high-z galaxies}",
      journal = {\mnras},
         year = 2017,
        month = nov,
       volume = {471},
       number = {4},
        pages = {4128-4143},
          doi = {10.1093/mnras/stx1792},
archivePrefix = {arXiv},
       eprint = {1707.04259},
 primaryClass = {astro-ph.GA},
       adsurl = {https://ui.adsabs.harvard.edu/abs/2017MNRAS.471.4128P}
}

@ARTICLE{Lupi2018,
       author = {{Lupi}, Alessandro and {Bovino}, Stefano and {Capelo}, Pedro R. and {Volonteri}, Marta and {Silk}, Joseph},
        title = "{The natural emergence of the correlation between H$_{2}$ and star formation rate surface densities in galaxy simulations}",
      journal = {\mnras},
         year = 2018,
        month = mar,
       volume = {474},
       number = {3},
        pages = {2884-2903},
          doi = {10.1093/mnras/stx2874},
archivePrefix = {arXiv},
       eprint = {1710.01315},
 primaryClass = {astro-ph.GA},
       adsurl = {https://ui.adsabs.harvard.edu/abs/2018MNRAS.474.2884L}
}

@ARTICLE{Hopkins2023,
       author = {{Hopkins}, Philip F. and {Wetzel}, Andrew and {Wheeler}, Coral and {Sanderson}, Robyn and {Grudi{\'c}}, Michael Y. and {Sameie}, Omid and {Boylan-Kolchin}, Michael and {Orr}, Matthew and {Ma}, Xiangcheng and {Faucher-Gigu{\`e}re}, Claude-Andr{\'e} and {Kere{\v{s}}}, Du{\v{s}}an and {Quataert}, Eliot and {Su}, Kung-Yi and {Moreno}, Jorge and {Feldmann}, Robert and {Bullock}, James S. and {Loebman}, Sarah R. and {Angl{\'e}s-Alc{\'a}zar}, Daniel and {Stern}, Jonathan and {Necib}, Lina and {Choban}, Caleb R. and {Hayward}, Christopher C.},
        title = "{FIRE-3: updated stellar evolution models, yields, and microphysics and fitting functions for applications in galaxy simulations}",
      journal = {\mnras},
         year = 2023,
        month = feb,
       volume = {519},
       number = {2},
        pages = {3154-3181},
          doi = {10.1093/mnras/stac3489},
archivePrefix = {arXiv},
       eprint = {2203.00040},
 primaryClass = {astro-ph.GA},
       adsurl = {https://ui.adsabs.harvard.edu/abs/2023MNRAS.519.3154H}
}

@ARTICLE{Konstandin2012,
       author = {{Konstandin}, L. and {Girichidis}, P. and {Federrath}, C. and {Klessen}, R.~S.},
        title = "{A New Density Variance-Mach Number Relation for Subsonic and Supersonic Isothermal Turbulence}",
      journal = {\apj},
         year = 2012,
        month = dec,
       volume = {761},
       number = {2},
          eid = {149},
        pages = {149},
          doi = {10.1088/0004-637X/761/2/149},
archivePrefix = {arXiv},
       eprint = {1206.4524},
 primaryClass = {astro-ph.SR},
       adsurl = {https://ui.adsabs.harvard.edu/abs/2012ApJ...761..149K}
}

@ARTICLE{Mocz2019,
       author = {{Mocz}, Philip and {Burkhart}, Blakesley},
        title = "{A Markov Model for Non-lognormal Density Distributions in Compressive Isothermal Turbulence}",
      journal = {\apjl},
         year = 2019,
        month = oct,
       volume = {884},
       number = {2},
          eid = {L35},
        pages = {L35},
          doi = {10.3847/2041-8213/ab48f6},
archivePrefix = {arXiv},
       eprint = {1908.00544},
 primaryClass = {astro-ph.GA},
       adsurl = {https://ui.adsabs.harvard.edu/abs/2019ApJ...884L..35M}
}

@ARTICLE{Nolan2015,
       author = {{Nolan}, C.~A. and {Federrath}, C. and {Sutherland}, R.~S.},
        title = "{The density variance-Mach number relation in isothermal and non-isothermal adiabatic turbulence}",
      journal = {\mnras},
         year = 2015,
        month = aug,
       volume = {451},
       number = {2},
        pages = {1380-1389},
          doi = {10.1093/mnras/stv1030},
archivePrefix = {arXiv},
       eprint = {1504.04370},
 primaryClass = {astro-ph.GA},
       adsurl = {https://ui.adsabs.harvard.edu/abs/2015MNRAS.451.1380N}
}

@ARTICLE{Scannapieco2026,
       author = {{Scannapieco}, Evan and {Br{\"u}ggen}, Marcus and {Grete}, Philipp and {Pan}, Liubin},
        title = "{An Improved Fit to the Density Fluctuations in Supersonic Isothermal Turbulence}",
      journal = {\apj},
         year = 2026,
        month = feb,
       volume = {998},
       number = {2},
          eid = {270},
        pages = {270},
          doi = {10.3847/1538-4357/ae394d},
archivePrefix = {arXiv},
       eprint = {2509.09811},
 primaryClass = {astro-ph.GA},
       adsurl = {https://ui.adsabs.harvard.edu/abs/2026ApJ...998..270S}
}

@ARTICLE{Federrath2008,
       author = {{Federrath}, Christoph and {Klessen}, Ralf S. and {Schmidt}, Wolfram},
        title = "{The Density Probability Distribution in Compressible Isothermal Turbulence: Solenoidal versus Compressive Forcing}",
      journal = {\apjl},
         year = 2008,
        month = dec,
       volume = {688},
       number = {2},
        pages = {L79},
          doi = {10.1086/595280},
archivePrefix = {arXiv},
       eprint = {0808.0605},
 primaryClass = {astro-ph},
       adsurl = {https://ui.adsabs.harvard.edu/abs/2008ApJ...688L..79F}
}

@ARTICLE{Passot1998,
       author = {{Passot}, Thierry and {V{\'a}zquez-Semadeni}, Enrique},
        title = "{Density probability distribution in one-dimensional polytropic gas dynamics}",
      journal = {\pre},
         year = 1998,
        month = oct,
       volume = {58},
       number = {4},
        pages = {4501-4510},
          doi = {10.1103/PhysRevE.58.4501},
archivePrefix = {arXiv},
       eprint = {physics/9802019},
 primaryClass = {physics.flu-dyn},
       adsurl = {https://ui.adsabs.harvard.edu/abs/1998PhRvE..58.4501P}
}

@ARTICLE{Padoan1997,
       author = {{Padoan}, Paolo and {Nordlund}, Ake and {Jones}, Bernard J.~T.},
        title = "{The universality of the stellar initial mass function}",
      journal = {\mnras},
         year = 1997,
        month = jun,
       volume = {288},
       number = {1},
        pages = {145-152},
          doi = {10.1093/mnras/288.1.145},
archivePrefix = {arXiv},
       eprint = {astro-ph/9703110},
 primaryClass = {astro-ph},
       adsurl = {https://ui.adsabs.harvard.edu/abs/1997MNRAS.288..145P}
}

@ARTICLE{Vazquez-Semadeni1994,
       author = {{Vazquez-Semadeni}, Enrique},
        title = "{Hierarchical Structure in Nearly Pressureless Flows as a Consequence of Self-similar Statistics}",
      journal = {\apj},
         year = 1994,
        month = mar,
       volume = {423},
        pages = {681},
          doi = {10.1086/173847},
       adsurl = {https://ui.adsabs.harvard.edu/abs/1994ApJ...423..681V}
}

@Article{Christensen2012,
  author        = {{Christensen}, C. and {Quinn}, T. and {Governato}, F. and {Stilp}, A. and {Shen}, S. and {Wadsley}, J.},
  journal       = {\mnras},
  title         = {{Implementing molecular hydrogen in hydrodynamic simulations of galaxy formation}},
  year          = {2012},
  month         = oct,
  pages         = {3058-3076},
  volume        = {425},
  adsurl        = {http://adsabs.harvard.edu/abs/2012MNRAS.425.3058C},
  archiveprefix = {arXiv},
  doi           = {10.1111/j.1365-2966.2012.21628.x},
  eprint        = {1205.5567},
}

@Book{Draine2011,
  author    = {{Draine}, B.~T.},
  publisher = {Princeton University Press},
  title     = {{Physics of the Interstellar and Intergalactic Medium}},
  year      = {2011},
  adsurl    = {http://adsabs.harvard.edu/abs/2011piim.book.....D},
  booktitle = {Physics of the Interstellar and Intergalactic Medium by Bruce T.~Draine.~Princeton University Press, 2011.~ISBN: 978-0-691-12214-4},
}

@Article{Federrath2010,
  author        = {{Federrath}, C. and {Banerjee}, R. and {Clark}, P.~C. and {Klessen}, R.~S.},
  journal       = {\apj},
  title         = {{Modeling Collapse and Accretion in Turbulent Gas Clouds: Implementation and Comparison of Sink Particles in AMR and SPH}},
  year          = {2010},
  month         = apr,
  pages         = {269-290},
  volume        = {713},
  adsurl        = {http://adsabs.harvard.edu/abs/2010ApJ...713..269F},
  archiveprefix = {arXiv},
  doi           = {10.1088/0004-637X/713/1/269},
  eprint        = {1001.4456},
  primaryclass  = {astro-ph.SR},
}

@Article{Gnedin2009,
  author        = {{Gnedin}, N.~Y. and {Tassis}, K. and {Kravtsov}, A.~V.},
  journal       = {\apj},
  title         = {{Modeling Molecular Hydrogen and Star Formation in Cosmological Simulations}},
  year          = {2009},
  month         = may,
  pages         = {55-67},
  volume        = {697},
  adsurl        = {http://adsabs.harvard.edu/abs/2009ApJ...697...55G},
  archiveprefix = {arXiv},
  doi           = {10.1088/0004-637X/697/1/55},
  eprint        = {0810.4148},
}

@ARTICLE{Hopkins2013b,
       author = {{Hopkins}, Philip F.},
        title = "{A model for (non-lognormal) density distributions in isothermal turbulence}",
      journal = {\mnras},
         year = 2013,
        month = apr,
       volume = {430},
       number = {3},
        pages = {1880-1891},
          doi = {10.1093/mnras/stt010},
archivePrefix = {arXiv},
       eprint = {1211.3119},
 primaryClass = {astro-ph.GA},
       adsurl = {https://ui.adsabs.harvard.edu/abs/2013MNRAS.430.1880H}
}

@Article{Krumholz2005,
  author    = {{Krumholz}, M.~R. and {McKee}, C.~F.},
  journal   = {\apj},
  title     = {{A General Theory of Turbulence-regulated Star Formation, from Spirals to Ultraluminous Infrared Galaxies}},
  year      = {2005},
  month     = sep,
  pages     = {250-268},
  volume    = {630},
  adsurl    = {http://adsabs.harvard.edu/abs/2005ApJ...630..250K},
  doi       = {10.1086/431734},
  eprint    = {astro-ph/0505177},
}

@Article{Kuhlen2012,
  author        = {{Kuhlen}, M. and {Krumholz}, M.~R. and {Madau}, P. and {Smith}, B.~D. and {Wise}, J.},
  journal       = {\apj},
  title         = {{Dwarf Galaxy Formation with H$_{2}$-regulated Star Formation}},
  year          = {2012},
  month         = apr,
  pages         = {36},
  volume        = {749},
  adsurl        = {http://adsabs.harvard.edu/abs/2012ApJ...749...36K},
  archiveprefix = {arXiv},
  doi           = {10.1088/0004-637X/749/1/36},
  eid           = {36},
  eprint        = {1105.2376},
}

@Article{Micic2012,
  author        = {{Micic}, M. and {Glover}, S.~C.~O. and {Federrath}, C. and {Klessen}, R.~S.},
  journal       = {\mnras},
  title         = {{Modelling H$_{2}$ formation in the turbulent interstellar medium: solenoidal versus compressive turbulent forcing}},
  year          = {2012},
  month         = apr,
  pages         = {2531-2542},
  volume        = {421},
  adsurl        = {http://adsabs.harvard.edu/abs/2012MNRAS.421.2531M},
  archiveprefix = {arXiv},
  doi           = {10.1111/j.1365-2966.2012.20477.x},
  eprint        = {1103.3056},
  primaryclass  = {astro-ph.SR},
}

@Article{Springel2005,
  author    = {{Springel}, V.},
  journal   = {\mnras},
  title     = {{The cosmological simulation code GADGET-2}},
  year      = {2005},
  month     = dec,
  pages     = {1105-1134},
  volume    = {364},
  adsurl    = {http://adsabs.harvard.edu/abs/2005MNRAS.364.1105S},
  doi       = {10.1111/j.1365-2966.2005.09655.x},
  eprint    = {arXiv:astro-ph/0505010},
}

@Article{Wolfire2003,
  author    = {{Wolfire}, M.~G. and {McKee}, C.~F. and {Hollenbach}, D. and {Tielens}, A.~G.~G.~M.},
  journal   = {\apj},
  title     = {{Neutral Atomic Phases of the Interstellar Medium in the Galaxy}},
  year      = {2003},
  month     = apr,
  pages     = {278-311},
  volume    = {587},
  adsurl    = {http://adsabs.harvard.edu/abs/2003ApJ...587..278W},
  doi       = {10.1086/368016},
  eprint    = {astro-ph/0207098},
}

@Article{Hopkins2015,
  author        = {{Hopkins}, P.~F.},
  journal       = {\mnras},
  title         = {{A new class of accurate, mesh-free hydrodynamic simulation methods}},
  year          = {2015},
  month         = jun,
  pages         = {53-110},
  volume        = {450},
  adsurl        = {http://adsabs.harvard.edu/abs/2015MNRAS.450...53H},
  archiveprefix = {arXiv},
  doi           = {10.1093/mnras/stv195},
  eprint        = {1409.7395},
}

@Article{Gaburov2011,
  author        = {{Gaburov}, E. and {Nitadori}, K.},
  journal       = {\mnras},
  title         = {{Astrophysical weighted particle magnetohydrodynamics}},
  year          = {2011},
  month         = jun,
  pages         = {129-154},
  volume        = {414},
  adsurl        = {http://adsabs.harvard.edu/abs/2011MNRAS.414..129G},
  archiveprefix = {arXiv},
  doi           = {10.1111/j.1365-2966.2011.18313.x},
  eprint        = {1006.4159},
  primaryclass  = {astro-ph.IM},
}

@Book{Tielens2005,
  author    = {{Tielens}, A.~G.~G.~M.},
  title     = {{The Physics and Chemistry of the Interstellar Medium}},
  year      = {2005},
  month     = aug,
  adsurl    = {http://adsabs.harvard.edu/abs/2005pcim.book.....T},
  booktitle = {The Physics and Chemistry of the Interstellar Medium, by A.~G.~G.~M.~Tielens, pp.~.~ISBN 0521826349.~Cambridge, UK: Cambridge University Press, 2005.},
}

@Article{Colbrook2017,
  author        = {{Colbrook}, M.~J. and {Ma}, X. and {Hopkins}, P.~F. and {Squire}, J.},
  journal       = {\mnras},
  title         = {{Scaling laws of passive-scalar diffusion in the interstellar medium}},
  year          = {2017},
  month         = may,
  pages         = {2421-2429},
  volume        = {467},
  adsurl        = {https://ui.adsabs.harvard.edu/abs/2017MNRAS.467.2421C},
  archiveprefix = {arXiv},
  doi           = {10.1093/mnras/stx261},
  eprint        = {1610.06590},
}

@Article{Larson1981,
  author   = {{Larson}, R.~B.},
  journal  = {\mnras},
  title    = {{Turbulence and star formation in molecular clouds.}},
  year     = {1981},
  month    = mar,
  pages    = {809-826},
  volume   = {194},
  adsurl   = {https://ui.adsabs.harvard.edu/abs/1981MNRAS.194..809L},
  doi      = {10.1093/mnras/194.4.809},
}

@Article{Hopkins2018a,
  author        = {{Hopkins}, Philip F. and {Wetzel}, Andrew and {Kere{\v{s}}}, Du{\v{s}}an and {Faucher-Gigu{\`e}re}, Claude-Andr{\'e} and {Quataert}, Eliot and {Boylan-Kolchin}, Michael and {Murray}, Norman and {Hayward}, Christopher C. and {Garrison-Kimmel}, Shea and {Hummels}, Cameron and {Feldmann}, Robert and {Torrey}, Paul and {Ma}, Xiangcheng and {Angl{\'e}s-Alc{\'a}zar}, Daniel and {Su}, Kung-Yi and {Orr}, Matthew and {Schmitz}, Denise and {Escala}, Ivanna and {Sanderson}, Robyn and {Grudi{\'c}}, Michael Y. and {Hafen}, Zachary and {Kim}, Ji-Hoon and {Fitts}, Alex and {Bullock}, James S. and {Wheeler}, Coral and {Chan}, T.~K. and {Elbert}, Oliver D. and {Narayanan}, Desika},
  journal       = {\mnras},
  title         = {{FIRE-2 simulations: physics versus numerics in galaxy formation}},
  year          = {2018},
  month         = oct,
  number        = {1},
  pages         = {800-863},
  volume        = {480},
  adsurl        = {https://ui.adsabs.harvard.edu/abs/2018MNRAS.480..800H},
  archiveprefix = {arXiv},
  doi           = {10.1093/mnras/sty1690},
  eprint        = {1702.06148},
  primaryclass  = {astro-ph.GA},
}
\bibliographystyle{aasjournal}

\appendix

\section{Numerical Convergence}\label{app:conv}

In this section, we present convergence tests where we compare simulations at $N = 64^3$, $128^3$, and $256^3$,
focusing on the clumping factors ($C_2$ and $C_3$) and the time evolution of the mixing efficiency.

\begin{figure*}
	\centering
	\includegraphics[width=0.9\linewidth]{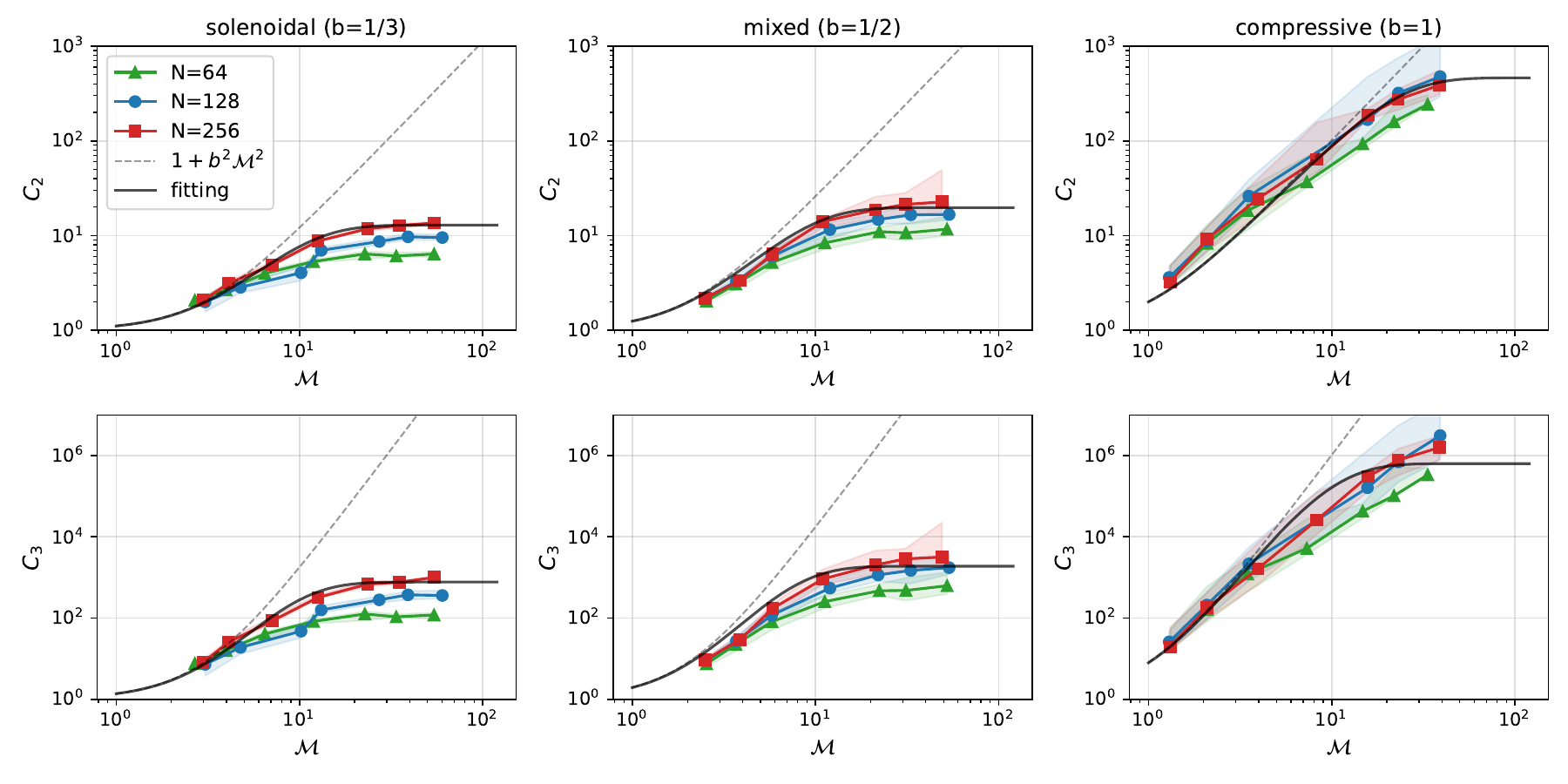}
	\caption{
    Convergence test for the two-body (top panels) and three-body (bottom panels) clumping factors as a function of the measured Mach number for solenoidal (left panels), mixed (middle panels), and compressive (right panels) driving modes at $N = 64^3$ (green), $128^3$ (blue), and $256^3$ (red), respectively.
    The solid curves show our fitting formula, while the dashed curves show the predictions from log-normal PDF theory.
    }
	\label{fig:cclump_convergence_C2C3}
\end{figure*}

Fig.~\ref{fig:cclump_convergence_C2C3} shows the convergence test for the two-body (top panels) and three-body (bottom panels) clumping factors as a function of the measured Mach number for solenoidal (left panels), mixed (middle panels), and compressive (right panels) driving modes at $N = 64^3$ (green), $128^3$ (blue), and $256^3$ (red), respectively.
The solid curves show our fitting formula, while the dashed curves show the predictions from log-normal PDF theory.

Broadly speaking, clumping factors are well converged at low $\mathcal{M}$ but they show systematic resolution dependence at high $\mathcal{M}$. 
Low-resolution runs underestimate clumping factors as they fail to capture the dense structures, leading to narrower PDFs and hence lower $C_n$.
The $N = 128^3$ and $N = 256^3$ runs converge within a factor of two and within the temporal scatter. 
In addition, their differences are significantly smaller than the deviation from the log-normal curve, suggesting that the saturation of $C_n$ at $\mathcal{M} \gtrsim 10$ is a physical effect rather than a resolution artifact.
Note that our fitting formula is a fit to the highest-resolution runs ($N=256^3$). 
Interestingly, the convergence improves with the fraction of compressive driving.
We interpret this as a result of Lagrangian codes being more capable of capturing coherent dense clumps caused by shock-driven compression due to their built-in adaptive resolution.
In contrast,
solenoidal driving produces thin, long filaments at intermediate densities, which are harder to resolve as the spatial resolution is not at its highest.

\begin{figure*}
	\centering
	\includegraphics[width=0.9\linewidth]{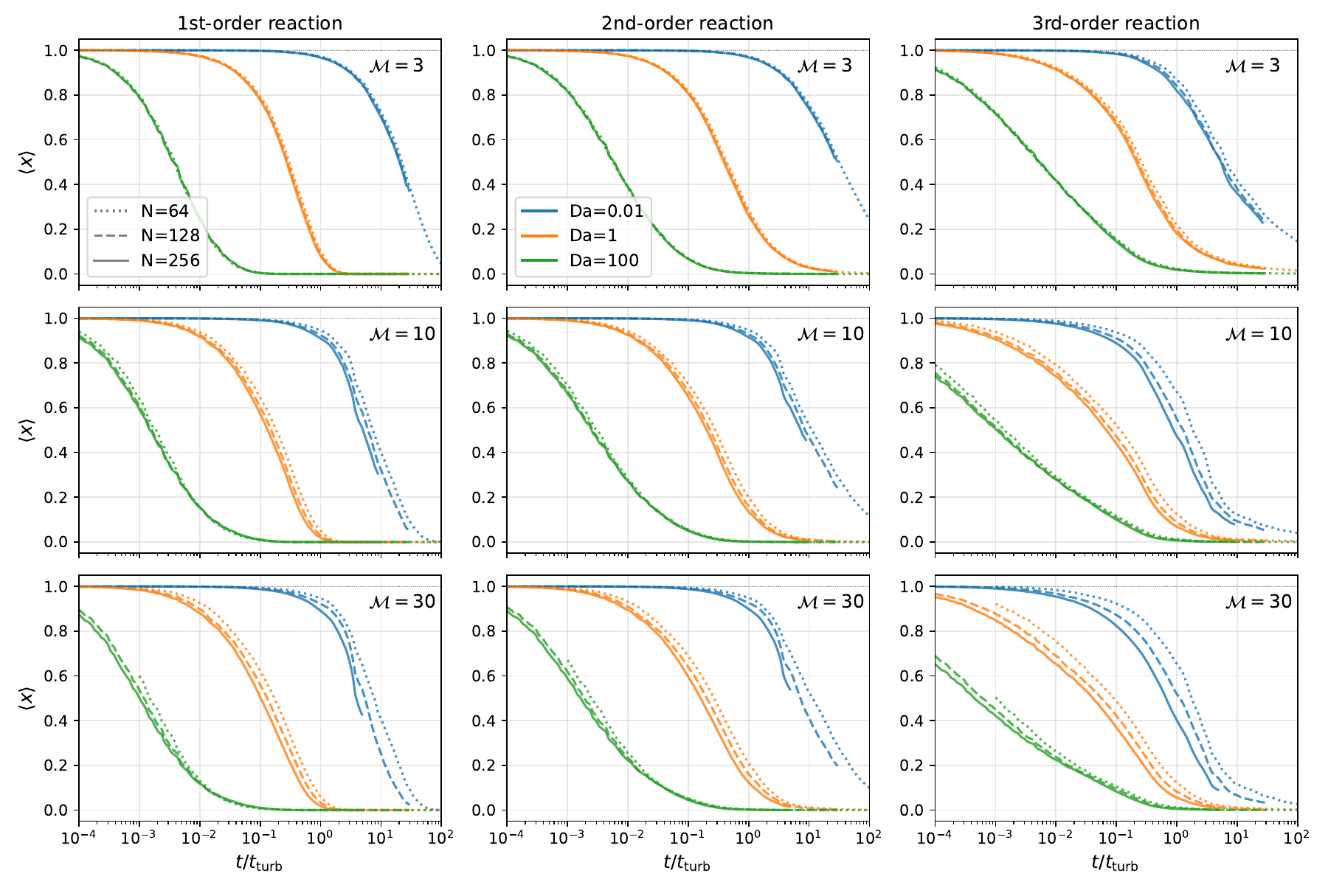}
	\caption{
    Convergence test for the time evolution of the mass-weighted average abundance ($\avM{x}$) in the simulations of $\mathcal{M} = 3$ (top), $\mathcal{M} = 10$ (middle), and $\mathcal{M} = 30$ (bottom) with Da $= 10^{2}$ (blue), $1$ (orange), $10^{-2}$ (green) for the 1st-, 2nd-, and 3rd-order reaction from left to right at resolution $N = 64^3$ (dotted), $128^3$ (dashed), and $256^3$ (solid).
    }
	\label{fig:abund_convergence_lin}
\end{figure*}

Fig.~\ref{fig:abund_convergence_lin} shows the convergence test for the time evolution of the mass-weighted average abundance $\avM{x}$ in the simulations of $\mathcal{M} = 3$ (top), $\mathcal{M} = 10$ (middle), and $\mathcal{M} = 30$ (bottom) with Da $= 10^{2}$ (blue), $1$ (orange), $10^{-2}$ (green) for the 1st-, 2nd-, and 3rd-order reaction from left to right at resolution $N = 64^3$ (dotted), $128^3$ (dashed), and $256^3$ (solid).

At low $\mathcal{M}$,
results show excellent convergence at all Da and for all reactions.
As $\mathcal{M}$ increases,
resolution effect becomes visible as the PDF broadens with $\mathcal{M}$ and becomes harder to resolve in low-resolution runs, suppressing the density contrast and leaving less room for the abundance-density anti-correlation.
Despite this trend with $\mathcal{M}$,
most results are reasonably well-converged.
The largest deviations appear in the 3rd-order reaction at high $\mathcal{M}$, with the $N = 128^3$ and $N = 256^3$ runs differing by about a factor of two.
This is because the 3rd-order reaction is highly sensitive to density, and so a small resolution effect on the density PDF can cause a notable difference on $\xi$ and thus $\avM{x}$.
We therefore caution that our $\xi$ for 3rd-order reactions can be slightly overestimated.

\section{Effect of subgrid abundance diffusion}\label{app:diff}

Lagrangian codes, such as {\sc Gizmo} used in this work, have no inherent numerical diffusion for passive scalars like chemical abundances.
This is in contrast to Eulerian codes where numerical diffusion can be regarded as a mechanism for sub-grid mixing.
As such, a potential concern of our Lagrangian simulations is that reactants, once depleted, have no mechanism to be replenished at the particle level,
which may lead to an artificial imprint on the density-abundance relation.
In this section, we investigate the effect of sub-grid diffusion of chemical abundances treated as passive scalars, following the metal diffusion module implemented in {\sc Gizmo} \citep{Colbrook2017, Hopkins2018a}.
We adopt a diffusion coefficient $a_q = 0.1$ based on the eddy diffusivity model calibrated in the \textit{a priori} test from \citet{Hu2020b}.

\begin{figure*}
	\centering
	\includegraphics[width=0.99\linewidth]{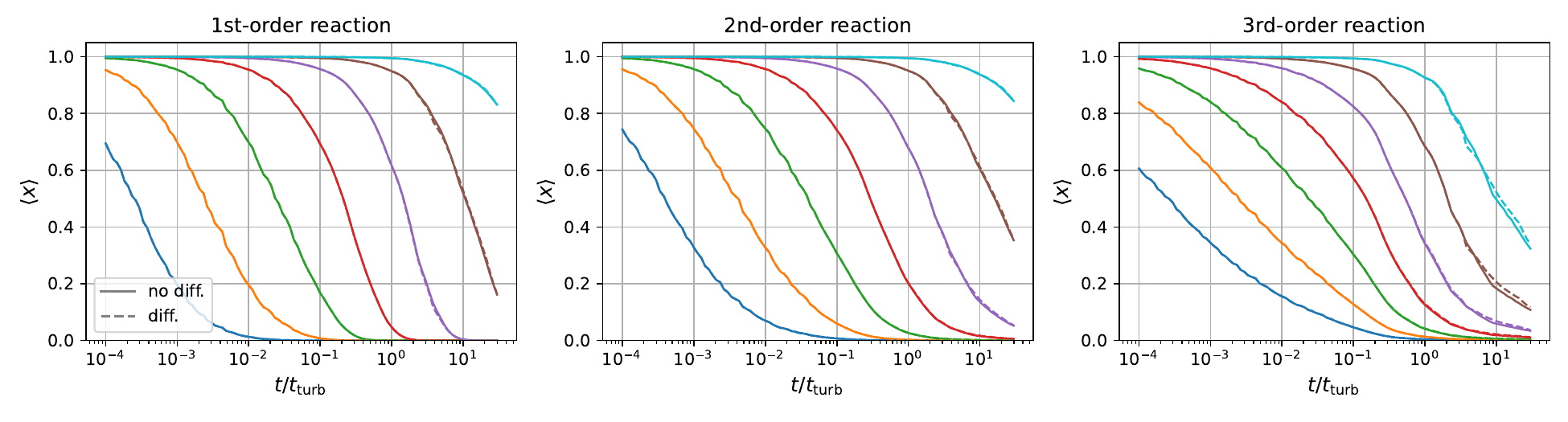}
	\caption{
    Time evolution of 
    the mass-weighted average abundance ($\avM{x}$) at $\mathcal{M} = 10$,
    with (dashed curves) and without (solid curves) sub-grid diffusion of chemical abundances.
    Different colors show abundances of different Da from $10^{-3}$--$10^{3}$.
    }
	\label{fig:compare_diff_abund_time}
\end{figure*}

Fig.~\ref{fig:compare_diff_abund_time} shows the time evolution of the mass-weighted average abundance ($\avM{x}$) at $\mathcal{M} = 10$, with (dashed curves) and without (solid curves) sub-grid diffusion of chemical abundances.
Different colors show abundances of different Da from $10^{-3}$--$10^{3}$.
Sub-grid abundance diffusion shows no visible effect in all cases,
so our results are robust against this numerical uncertainty.
This indicates that turbulent transport of abundance is dominated by the resolved flows rather than by sub-grid mixing.
In other words,
in a well-mixed case (low Da),
each gas particle cycles through the same density PDF and thus accumulates a similar net depletion of reactants.


\end{document}